\documentclass[%
 reprint,
superscriptaddress,
 amsmath,amssymb,
 aps,
pra,
floatfix,
]{revtex4-2}

\usepackage{graphicx}
\usepackage{xcolor}
\usepackage{dcolumn}
\usepackage{bm}
\usepackage{braket}
\usepackage{stmaryrd}
\usepackage[caption=false]{subfig}
\usepackage{here}
\usepackage{placeins}
\usepackage{hyperref}
\hypersetup{
    unicode=false,          
    pdftoolbar=true,        
    pdfmenubar=true,        
    pdffitwindow=false,     
    pdfstartview={FitH},    
    pdfkeywords={keyword1} {key2} {key3}, 
    pdfnewwindow=true,      
    colorlinks=true,       
    linkcolor=blue,          
    citecolor=blue,        
    filecolor=magenta,      
    urlcolor=blue,
}

\begin{document}

\title{Dense packing of the surface code: code deformation procedures and hook-error-avoiding gate scheduling}
\author{Kohei Fujiu}
\affiliation{Graduate School of Science and Technology, Keio University, Yokohama, Kanagawa, 223-8522, Japan}

\author{Shota Nagayama}
\affiliation{Graduate School of Media Design, Keio University, Yokohama, Kanagawa, 223-0061, Japan}

\author{Shin Nishio}
\affiliation{Graduate School of Science and Technology, Keio University, Yokohama, Kanagawa, 223-8522, Japan}
\affiliation{Department of Physics \& Astronomy, University College London, London, WC1E 6BT, United Kingdom}

\author{Hideaki Kawaguchi}
\affiliation{Graduate School of Science and Technology, Keio University, Yokohama, Kanagawa, 223-8522, Japan}

\author{Takahiko Satoh}
\affiliation{Faculty of Science and Technology, Keio University, Yokohama, Kanagawa, 223-8522, Japan}

\date{\today}

\begin{abstract}
The surface code is one of the leading quantum error correction codes for realizing large-scale fault-tolerant quantum computing (FTQC).
One major challenge in realizing surface-code-based FTQC is the extremely large number of qubits required. To mitigate this problem, fusing multiple codewords of the surface code into a densely packed configuration has been proposed.
It is known that by using dense packing, the number of physical qubits required per logical qubit can be reduced to approximately three-fourths compared to simply placing surface-code patches side by side.
Despite its potential, concrete deformation procedures and quantitative error-rate analyses have remained largely unexplored.
In this work, we present a detailed code-deformation procedure that transforms multiple standard surface code patches into a densely packed, connected configuration, along with a conceptual microarchitecture to utilize this dense packing. We also propose a CNOT gate-scheduling for stabilizer measurement circuits that suppresses hook errors in the densely packed surface code. 
We performed circuit-level Monte Carlo noise simulation of densely packed surface codes using this gate scheduling.
The numerical results demonstrate that as the code distance of the densely packed surface code increases and the physical error rate decreases, the logical error rate of the densely packed surface code becomes lower than that of the standard surface code.
Furthermore, we find that only when employing hook-error-avoiding syndrome extraction can the densely packed surface code achieve a lower logical error rate than the standard surface code, while simultaneously reducing the space overhead. 
\end{abstract}

\maketitle

\section{Introduction}
\label{sec:intro}
Quantum computer is expected to solve many important classically intractable problems~\cite{shor1999polynomial, gidney2021factor, bauer2020quantum}. However, the performance of a quantum computer is fundamentally constrained by the coherence times of its physical qubits~\cite{divincenzo2000physical}. To reliably store and manipulate quantum information in the presence of noise, quantum error-correcting codes, which encode logical quantum information onto redundant physical qubits, have been proposed~\cite{shor1995scheme, preskill1998reliable, terhal2015quantum}.

Topological quantum error-correcting codes~\cite{kitaev2003fault, terhal2015quantum} have attracted particular attention, as they require only the measurement of spatially local operators. This property makes them compatible with the two-dimensional solid-state qubit architectures such as superconducting qubits~\cite{google2025quantum}, and spin qubits~\cite{xue2022quantum}.
Within this class, the surface code~\cite{bravyi1998quantum, fowler2012surface} stands out as one of the most promising codes. For fault-tolerant quantum computation with the surface code in two-dimensional architectures, logical gate implementation techniques such as lattice surgery~\cite{horsman2012surface} and defect braiding~\cite{fowler2012surface} have been proposed.
Lattice surgery has attracted particular attention because it enables computation with much lower resource overhead~\cite{fowler2018low, litinski2018lattice, litinski2019game}.
In lattice surgery-based computation, multiple patches of surface code are arranged and manipulated~\cite{horsman2012surface, fowler2018low}. Here, a patch refers to an independent codeword of surface code with boundaries, encoding a single logical qubit. In this work, we specifically treat a patch as a rotated surface code~\cite{bombin2007optimal, anderson2013homological, tomita2014low, o2025compare}.

One major drawback of the surface code is its significant demand for a large number of physical qubits per logical qubit. The number of physical qubits required to construct a single patch of the surface code increases proportionally to the square of the code distance. To alleviate this issue, several schemes have been investigated to reduce the physical-qubit overhead per a codeword of the surface code~\cite{hastings2014reduced, nagayama2017state, gidney2021factor, gidney2025yoked}.

In this work, we focus on dense packing~\cite{gidney2021factor}, which fuses multiple patches of the surface code and stores their codewords at higher density. This method has been proposed to reduce the number of physical qubits required per logical qubit to approximately three-quarters. This method could contribute to alleviating the physical qubit requirements in surface code–based fault-tolerant quantum computation.
The dense packing is expected to potentially introduce a memory hierarchy in fault-tolerant quantum computing~\cite{copsey2002memory, Oskin_2002, Thaker, kobori2025lsqca} and reduce its spatial overhead. 
However, methods for transitioning from the state of a standard surface code to a densely packed state, as well as the logical error rates in a densely packed configuration, have not been clarified~\cite{gidney2021factor}. In this work, we propose a code deformation procedure that transforms standard patches into a densely packed state. We further designed a CNOT gate scheduling scheme for syndrome measurements tailored to this densely packed configuration, and validated its effectiveness through logical error rate simulations.

The structure of this paper is as follows. First, Section~\ref{sec:pre} provides an overview of the fundamental concepts of the surface code as preliminary knowledge, followed by an explanation of the code deformation necessary to achieve dense packing. Section~\ref{sec:reduce} discusses the upper limit of physical qubits that can be reduced in densely packed surface codes. We also propose a conceptual microarchitecture that employs dense packing for space-efficient fault-tolerant quantum computing. 
Section~\ref{sec:transforming} describes a code deformation procedure that transforms standard surface code patches into a densely packed configuration.
Section~\ref{sec:gate} provides CNOT gate-scheduling for stabilizer measurement circuits in densely packed surface code, which suppresses hook errors.
Finally, Section~\ref{sec:simulation} presents simulation results of the logical error rate in the densely packed state. We found that densely packed surface code demonstrates logical error rates comparable to or lower than those of standard surface code.

\section{background}
\label{sec:pre}
This section explains the basic aspects of surface codes and how they can be deformed.

\subsection{\label{sec:level2}Surface Codes}
Reliable quantum computation requires protecting fragile quantum states from errors. This protection can be achieved by encoding logical qubits onto redundant physical qubits with quantum error-correcting (QEC) codes. Many QEC codes, including the surface codes, are constructed on the basis of the stabilizer formalism~\cite{gottesman1997stabilizer}. By performing syndrome measurements of stabilizer generators, errors can be detected without destroying the encoded logical information.

Among the various QEC codes, the surface code~\cite{bravyi1998quantum, kitaev2003fault, bombin2007optimal, anderson2013homological, tomita2014low, o2025compare} stands out as a leading candidate due to its high threshold error rate~\cite{wang2009threshold} and compatibility with two-dimensional architectures requiring only nearest-neighbor interactions. As a result, many fault-tolerant quantum computing architectures~\cite{byun2022xqsim, duckering2020virtualized, ueno2022qulatis} are based on the surface code.

\begin{figure}[htbp]
\includegraphics[width=0.35\textwidth]{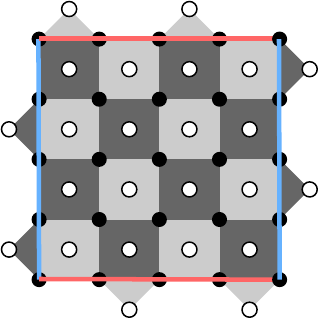}
\caption{\label{fig:bsurface} A $\llbracket 25, 1, 5 \rrbracket$ surface code patch. The black dots represent the data qubits of the patch, while the white dots denote the measurement qubits used for syndrome measurements. The dark-shaded regions and light-shaded regions represent the $X^{\bigotimes 4}(X^{\bigotimes 2})$ and $Z^{\bigotimes 4}(Z^{\bigotimes 2})$ stabilizer generators, respectively. Here, $X^{\bigotimes 4}$ and $Z^{\bigotimes 4}$ denote the square stabilizer generators acting on four data qubits, and $X^{\bigotimes 2}$ and $Z^{\bigotimes 2}$ denote the boundary stabilizer generators acting on  two data qubits at the edges of the patch. The red lines on the patch denote $X$ boundaries, and the blue lines denote $Z$ boundaries.}
\end{figure}

\begin{figure}
    \centering
    \subfloat[][\centering]{
    \includegraphics[keepaspectratio, width=0.3\linewidth]{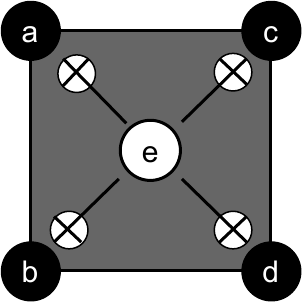}
    }
    \subfloat[][\centering]{
    \includegraphics[keepaspectratio, width=0.65\linewidth]{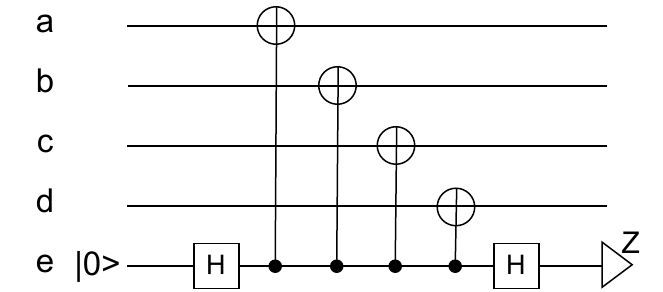}
    }
    \hfill
    \subfloat[][\centering]{
    \includegraphics[keepaspectratio, width=0.3\linewidth]{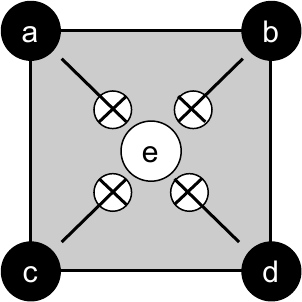}
    }
    \subfloat[][\centering]{
    \includegraphics[keepaspectratio, width=0.65\linewidth]{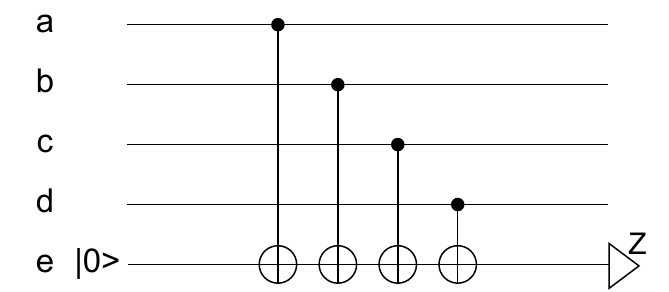}
    }
    \caption{(a) $X$-type stabilizer of a surface code which corresponds to each dark shaded region in Fig.~\ref{fig:bsurface}. It consists of supports of the $X^{\bigotimes 4}$ stabilizer as black dots and a measurement qubit as a white dot in the same way as Fig.~\ref{fig:bsurface}. Four lines with a $\oplus$ sign depicted in the figure correspond to commutative $CX$ gates necessary for syndrome measurement. (b) A circuit implementation of syndrome measurement for a $X^{\bigotimes 4}$ stabilizer. (c) $Z$-type stabilizer of the surface code, which corresponds to each light shaded region in Fig.~\ref{fig:bsurface}. (d) A circuit implementation of syndrome measurement for a $Z^{\bigotimes 4}$ stabilizer.}
\label{fig:X-type_stabilizer}
\end{figure}

Fig.~\ref{fig:bsurface} shows a patch of the surface code. In this configuration, physical qubits are arranged on a two-dimensional lattice. 
The black dots denote physical data qubits, while the white dots represent the auxiliary qubits used for syndrome measurements, which we refer to as measurement qubits. The regions delineated by squares and triangles represent stabilizer generators. Here, dark-shaded areas correspond to $X$-type stabilizers and light-shaded areas correspond to $Z$-type stabilizers. Errors on data qubits are detected through syndrome measurements of the stabilizer generators. The red and blue lines are referred to as the $X$ and $Z$ boundaries, respectively. A sequence of $X/Z$ gate operations on the physical qubits, along with $X/Z$ boundaries, corresponds to a logical $X/Z$ gate on the surface code. The code distance $d$ of a surface code is defined as the minimum weight of a nontrivial logical operator.

Fig.~\ref{fig:X-type_stabilizer}(a) shows the single $X$-type stabilizer generator excerpted, illustrating required two-qubit gates between data qubits and a measurement qubit for the syndrome measurement to perform the syndrome extraction. Fig.~\ref{fig:X-type_stabilizer}(b) depicts the exact quantum circuits for the syndrome extraction. Fig.~\ref{fig:X-type_stabilizer}(c) and (d) similarly describe the $Z$-type stabilizer. When an odd number of $Z$ errors occur on this $X$-type stabilizer, the syndrome obtained from this stabilizer flips. Errors on data qubits are corrected using the outcomes of syndrome measurements of the stabilizer generators. However, these syndrome measurements themselves are subject to noise. To reliably perform error correction, it is therefore necessary to repeat the syndrome measurements multiple times~\cite{fowler2012surface}. A complete cycle in which all stabilizers are measured once is referred to as a round, and typically $d$ rounds of measurements are performed where $d$ is equal to the code distance.  This ensures that errors, including measurement errors, can be corrected, thereby achieving protection with a distance $d$ in both time and space.

\begin{figure}[htbp]
\includegraphics[width=0.45\textwidth]{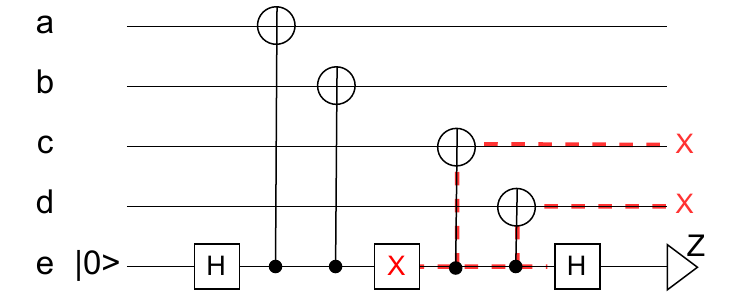}
\caption{\label{fig:hook_error} An example of the hook error in a syndrome extraction circuit. An X error on the measurement qubit propagates to the two data qubits c and d.}
\end{figure}

In stabilizer syndrome measurement circuits, errors occurring on measurement qubits can propagate to multiple data qubits. Fig.~\ref{fig:hook_error} illustrates an example where a single $X$ error on the measurement qubit during the measurement of an 
$X^{\otimes{4}}$-stabilizer propagates as $X$ errors on two data qubits. Such errors are referred to as hook errors~\cite{tomita2014low}. For $Z$-type stabilizers, an analogous argument holds with $X$ and $Z$ interchanged. 

\begin{figure}[htbp]
\includegraphics[width=0.35\textwidth]{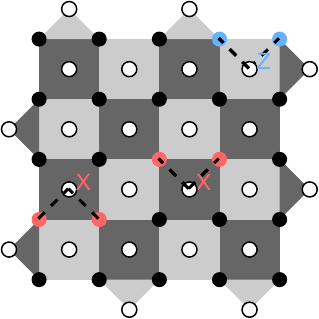}
\caption{\label{fig:hook_error_on_patch} As shown in Fig.~\ref{fig:hook_error}, errors on measurement qubits propagate to two data qubits. The $X$-type hook errors propagate along the logical $X$ operator, while the $Z$-type hook error propagate perpendicular to the logical $Z$ operator.}
\end{figure}

Fig.~\ref{fig:hook_error_on_patch} illustrates how hook errors in the patch extend the length of error chains. In this case, $X$ errors occurring on the measurement qubits of the $X$-type stabilizer generators propagate to the data qubits, turning two physical errors into four physical errors. Because the errors align with the logical $X$ operator in the patch, a logical $X$ error can occur with just two physical errors, despite the original code distance being five. Additionally, a $Z$ error occurring on the measurement qubit of the $Z$-type stabilizer generator propagates to the data qubits, resulting in one physical error becoming two physical errors on the data qubits. However, the two propagated $Z$ errors are orthogonal to the logical $Z$ operator and do not increase the logical error rate of the patch. 

Since hook errors may cause error propagations that increase the logical error rate, two-qubit gate scheduling must be determined with these propagations in mind~\cite{litinski2018lattice, chamberland2022circuit, gidney2023cleaner, o2025compare}. 
The propagation direction of hook errors is determined by the scheduling of the two-qubit gates. By determining appropriate gate scheduling so that $X/Z$-type hook errors occur perpendicular to the direction of the $X/Z$ logical operators on the patch, the minimum number of physical errors required to cause a logical error can be kept unchanged even when hook errors occur. 
Section~\ref{sec:gate} presents a detailed discussion of gate scheduling.

\subsection{How to deform Surface Codes}
\begin{figure}[htbp]
    \centering
    \subfloat[][A surface code patch and physical qubits initialized to the $\ket{0}$ state.]{
    \includegraphics[keepaspectratio, width=0.45\linewidth]{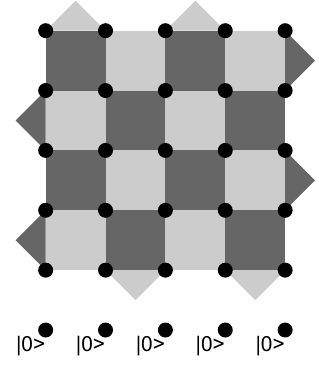}
    }
    \hfill
    \subfloat[][By introducing additional stabilizers to the surface code patch, it can be extended after $d$ rounds of sysndrome measurement.]{
    \includegraphics[keepaspectratio, width=0.45\linewidth]{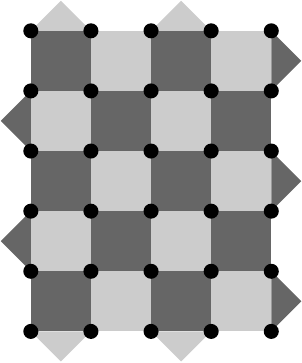}
    }

    \subfloat[][By measuring the data qubits on the surface code in the $Z$ basis, it can become shorter.]{
    \includegraphics[keepaspectratio, width=0.45\linewidth]{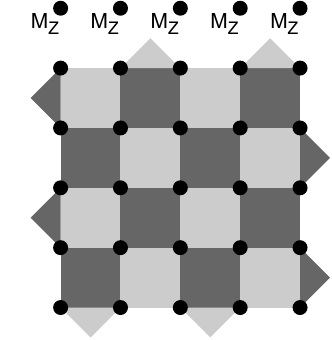}
    }
    
    \caption{Code deformation procedures to extend and contract a surface code patch.}
\label{fig:Code Deformation}
\end{figure}

The configuration of a code can be altered via code deformation~\cite{bombin2009quantum}. Code deformation is a technique that modifies the generating set of measured stabilizers to transition the code from one state to another. In Fig.~\ref{fig:Code Deformation}, code deformation is used to extend and contract a surface code patch. Fig.~\ref{fig:Code Deformation}(a) shows a single patch of the surface code together with physical qubits, which are initialized in the $\ket{0}$ state. In Fig.~\ref{fig:Code Deformation}(b), stabilizer syndrome measurements are performed including the physical qubits initialized in (a). After $d$ rounds of measurements, the patch is effectively expanded. Also, the patch can be reduced by measuring its physical qubits. For instance, in Fig.~\ref{fig:Code Deformation}(c), measuring the data qubits along the $X$-boundary in the $Z$ basis shortens the length of the $Z$-boundary. In Section~\ref{sec:transforming}, we demonstrate that this method enables multiple surface code patches to be transformed into a more densely packed configuration and restored.

\section{Reducing the space overhead of Surface Code}
\label{sec:reduce}
This section discusses the space overhead of surface codes that can be reduced by dense packing and a conceptual microarchitecture utilizing dense packing.

\subsection{Reducing Spatial Overhead via Dense Packing}
\begin{figure}[htbp]
\includegraphics[width=0.4\textwidth]{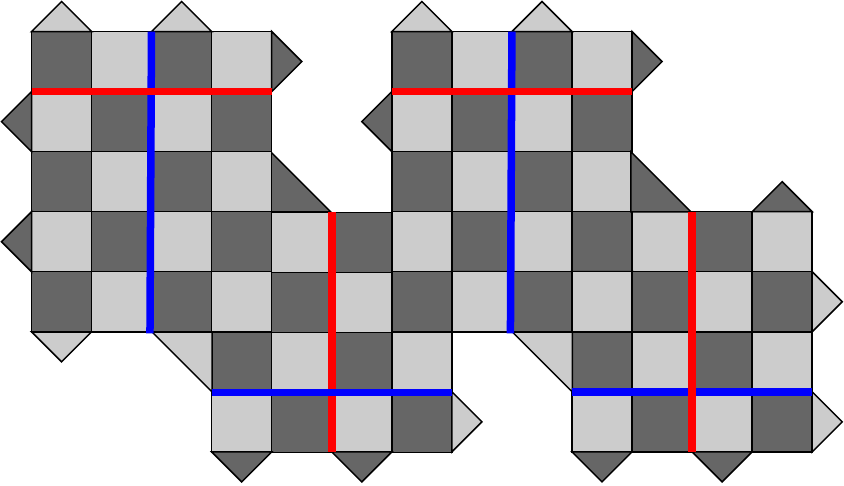}
\caption{\label{fig:gidney_dense3} An example of densely packed~\cite{gidney2021factor} surface code codewords. Four logical qubits are packed in this case, thereby realizing a more efficient encoding that requires a reduced area, resulting in lower physical-qubit overhead. The red and blue lines correspond to the logical $X$ and $Z$ operators of the logical qubits, respectively.}
\end{figure}

In the previous study~\cite{gidney2021factor}, it has been proposed to deform multiple surface codes into a single high-density code as illustrated in Fig.~\ref{fig:gidney_dense3}. In this configuration, the logical operators $X$ and $Z$ are represented as red and blue lines, respectively. Although multiple surface codes are fused into a single state, the original logical states are preserved. This configuration allows the logical state to be maintained with fewer physical qubits than in the standard surface-code layout. We refer to this technique as dense packing.
In contrast to this technique, in this paper, we refer to the conventional, non-densely packed surface code patch as the \emph{Standalone surface code patch}, to emphasize that it functions as an independent single logical qubit patch with a separated layout.

Although the prior study has suggested that dense packing reduces the number of qubits used to approximately three-fourths of the original count, the details of this reduction were not explicitly clarified. 
This also applies to the allocation strategy of densely packed surface codes. In addition to the horizontal fusion forming a row as shown in Fig.~\ref{fig:gidney_dense3}, vertical fusion can also be considered.
In this section, we examine the conditions under which the allocation strategy of dense packing becomes effective and show the reduction rate of the space overhead of surface codes under these conditions. 

In the previous study, the patch lengths of the upper and lower surface codes were different. However, in this discussion, it is assumed that the patch lengths are the same in both the $X$ and $Z$ directions, and that all surface codes have identical patch lengths. It is assumed that the patch length of the surface codes is an odd number. 

\begin{figure}[htbp]
\includegraphics[width=0.4\textwidth]{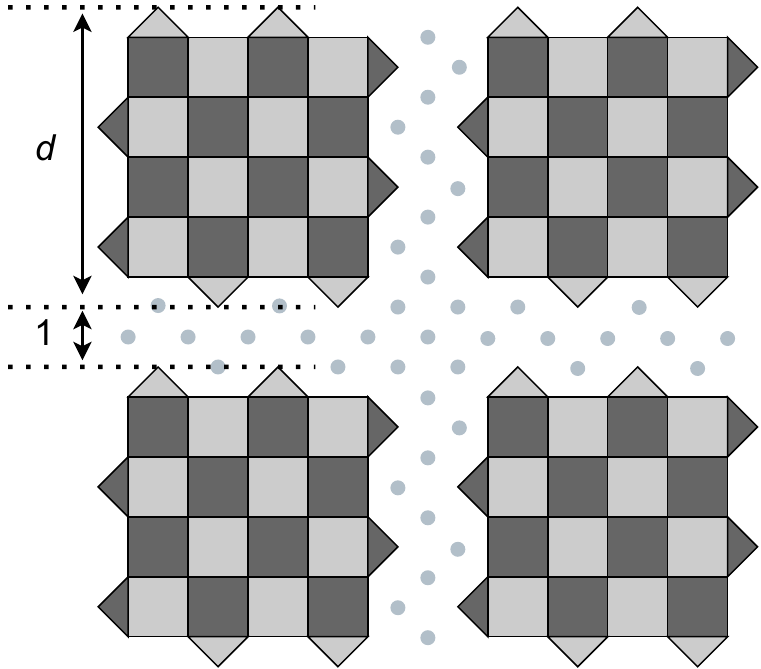}
\caption{\label{fig:normal_dense} A layout of standalone surface code patches without dense packing. The gray dots represent physical qubits. These physical qubits are not accessed for quantum computation.}
\end{figure}

As a measure of the physical qubit requirement per logical qubit, we use the space occupied by a single logical qubit rather than the number of physical qubits required for its encoding. 
This is because, when surface code patches are arranged on a square lattice, there arise physical qubits that cannot be accessed for computation, even though they are not used as data qubits or measurement qubits on the surface code. This is shown in Fig.~\ref{fig:normal_dense}. Therefore, in the following discussion, we evaluate the space overhead of the surface code required per logical qubit.

When standalone surface code patches are arranged, their layout appears as shown in Fig.~\ref{fig:normal_dense}, where $n_h$ codes are placed vertically, and $n_w$ codes are placed horizontally. The total area occupied by these $n_hn_w$ surface codes is given by:
\begin{align}
&\{n_hd+(n_h-1)\}\{n_wd+(n_w-1)\}\notag \\= \quad &n_hn_w(d+1)^2-(n_h+n_w)(d+1)+1.
\end{align}
Consequently, the area occupied per logical qubit is:
\begin{align}
(d+1)^2-\frac{(n_h+n_w)(d+1)}{n_hn_w}+\frac{1}{n_hn_w}.
\end{align}

\begin{figure}[htbp]
\includegraphics[width=0.48\textwidth]{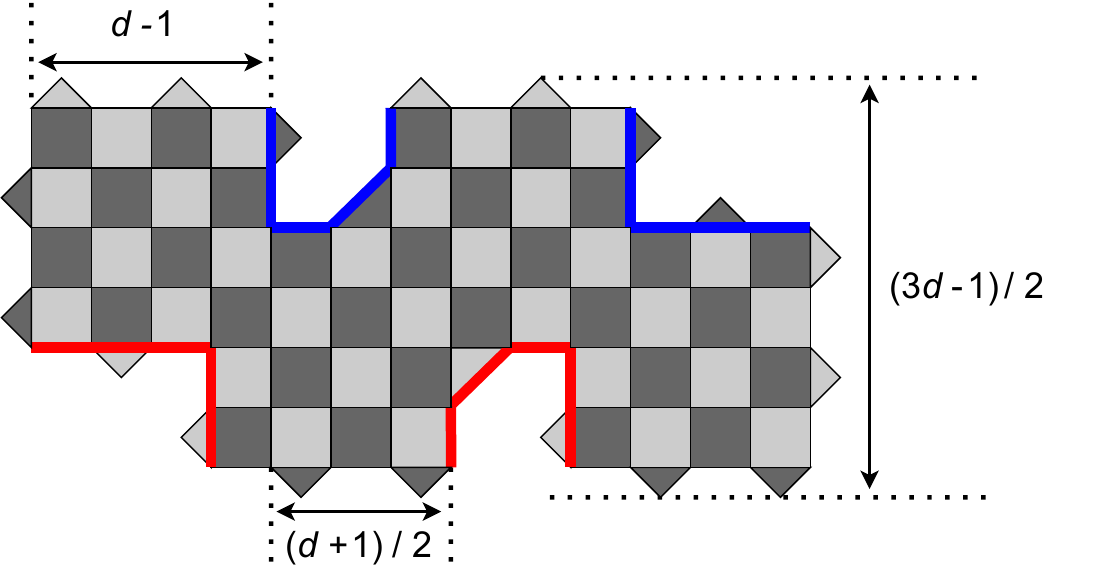}
\caption{\label{fig:dense_num2} Densely packed surface codes. The individual code distances are kept as close to $d$ as possible, and the distances of logical operators across multiple codes, as shown by the red and blue lines, are kept $d$ or more. The red and blue lines correspond to the $X$-type and $Z$-type multi-qubit logical operators, respectively.}
\end{figure}

On the other hand, when the surface codes are arranged with dense packing while maintaining the same patch length, their layout changes to that shown in Fig.~\ref{fig:dense_num2}. In this case, the total area occupied by $n$ surface codes is:
\begin{align}
    &\frac{3d-1}{2}\{(d-1)+(n-1)\frac{d+1}{2}+1\}\notag \\= \quad&\frac{(3d-1)(d+1)n}{4}+\frac{(3d-1)(d-1)}{4}.
\end{align}
Assuming $n \gg d^2$, the area occupied per logical qubit becomes:
\begin{align}
    &\frac{(3d-1)(d+1)}{4}+\frac{(3d-1)(d-1)}{4n}\notag \\ \approx \quad&\frac{(3d-1)(d+1)}{4}.
\end{align}

By comparing this result to the first case, it can be confirmed that as $n$ becomes significantly larger than $d^2$, the area occupied per logical qubit asymptotically reduces to 3/4 of the original value.

\begin{figure}[htbp]
\includegraphics[width=0.4\textwidth]{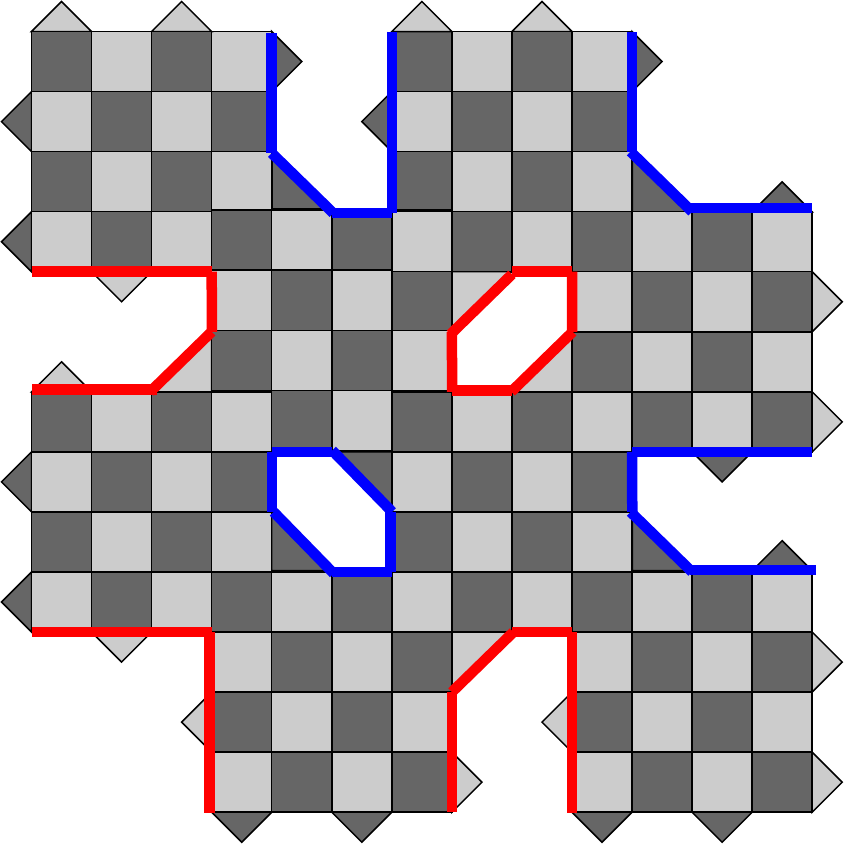}
\caption{\label{fig:possible} Example of Dense packing extension. In addition to using dense packing in the horizontal direction, it is possible to use it in the vertical direction as well. The individual code distances are kept as close to $d$ as possible, and the distances of logical operators across multiple codes, as shown by the red and blue lines, are kept $d$ or more. The red and blue lines correspond to the $X$-type and $Z$-type multi-qubit logical operators, respectively.}
\end{figure}

By extending the dense packing arrangement shown in Fig.~\ref{fig:dense_num2} vertically, it is conceivable to further increase the number of surface codes packed densely. For example, the configuration illustrated in Fig.~\ref{fig:possible} can be considered. If the arrangement in Fig.~\ref{fig:possible} consists of $n_w$ surface codes per row and $n_h$ such rows stacked vertically, the total area occupied by the densely packed surface codes is given by:

\begin{align}
&\{\frac{3d-3}{2}(n_h-1)+\frac{3d-1}{2}+1\}\\
& \quad \times \{(d-1)+\frac{d+1}{2}(n_w-1)+1\}\notag\\
=\quad &\frac{1}{4}\{(3d-3)(d+1)n_hn_w +(3d-3)(d-1)n_h\notag \\& \quad +4(d+1)n_w+4(d-1)\}.
\end{align}
The area occupied per logical qubit is then expressed as:
\begin{align}
    \frac{1}{4}\{(3d-3)(d+1)&+\frac{(3d-3)(d-1)}{n_w}\notag\\  &+\frac{4(d+1)}{n_h}+\frac{4(d-1)}{n_hn_w}\}.
\end{align}
As $n_w$ becomes larger, the area used relative to that of a standalone surface code asymptotically approaches $3/4$. 
Therefore, it is preferable to make $n_w$ as large as possible.
This preference arises because vertical dense packing introduces defects that span multiple surface code patches, creating logical structures whose code distance must be preserved. 

When used as quantum storage, accessing logical qubits located at the center of the structure becomes challenging. Furthermore, due to defects in the densely packed surface code, gate scheduling is expected to be more complex than the case with Fig.~\ref{fig:dense_num2}. The exact procedure will be described later, only for the linear configuration in Section~\ref{sec:gate}. For practical quantum computing, it may be more efficient to maintain multiple instances of the configuration shown in Fig.~\ref{fig:dense_num2}. We also note that, although denser configurations based on dense packing than that shown in Fig.~\ref{fig:possible} may exist, we focus on the approach using Fig.~\ref{fig:dense_num2} for the same practical considerations discussed above.

\subsection{A Conceptual Microarchitecture with Hierarchical Memory}
In large-scale fault-tolerant quantum computing, many logical qubits act as long-term memory rather than active computation units. Treating all patches equally leads to inefficient use of spatial resources. By employing densely packed surface codes as a memory layer, we can realize a hierarchical quantum memory that trades off access time for spatial efficiency.

Here, we propose one possible conceptual design of a microarchitecture that introduces a memory hierarchy by employing both dense-packed and standalone surface codes. This microarchitecture is divided into two regions: a computational region and a memory region, inspired by the Load/Store architecture~\cite{kobori2025lsqca}. The computational region adopts the standalone configuration to achieve highly parallel time-efficient computation, while the memory region utilizes multiple rows of densely packed surface codes to provide space-efficient storage for patches that are not frequently accessed.

\begin{figure}[htbp]
    \centering
    \subfloat[][\centering]{
    \includegraphics[keepaspectratio, width=0.47\linewidth]{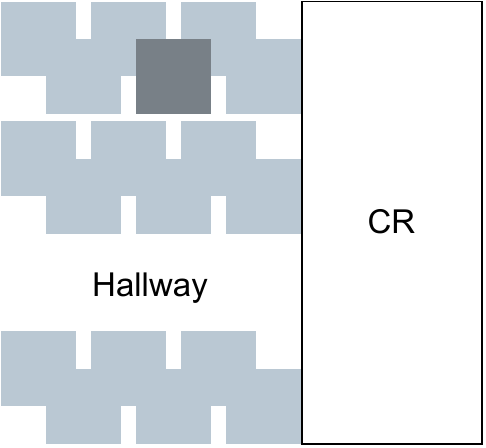}
    }
    \hfill
    \subfloat[][\centering]{
    \includegraphics[keepaspectratio, width=0.47\linewidth]{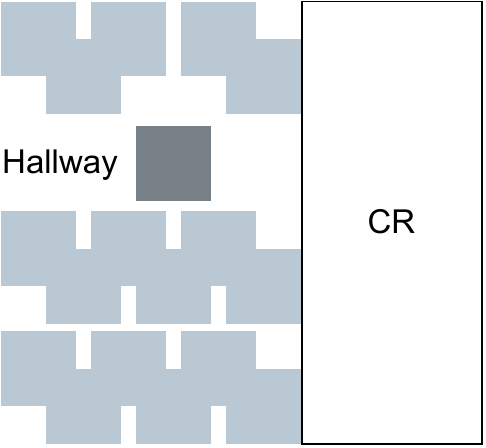}
    }
    \caption{A layout for a microarchitecture with hierarchical memory. Densely packed surface codes are used to provide space-efficient storage for infrequently accessed logical qubits. (a) The logical patch that the computational region (CR) needs to access is depicted as a black square, but other rows are obstructing it. (b) By moving the second row downward, the logical patch faces the hallway, allowing access from the computational region.}
\label{fig:floorplan}
\end{figure}

Fig.~\ref{fig:floorplan} shows the layout of the microarchitecture. The light gray areas represent densely packed surface codes and a standalone surface code patch in the memory region. In Fig.~\ref{fig:floorplan}(a), densely packed rows are arranged, and also a row of empty qubits --- referred to as a  \emph{hallway} --- is provided to enable selective access to specific logical qubits from the computational region. 

When attempting to access the logical information stored in a row that is not facing the hallway (for example, the topmost row, illustrated in dark gray in Fig.~\ref{fig:floorplan}), the other row (the second row in this example) obstructs direct access to the first row’s logical information. In such a case, one can shift the entire second row downward as shown in  Fig.~\ref{fig:floorplan}(b) with $2d$ rounds plus two parallel physical qubits measurement steps. The code deformation procedure for the movement of the entire row of densely packed surface codes is provided in Appendix~\ref{sec:app_movement}. After this shift, the first row faces the hallway, and the desired patch is separated from its neighbors. This movement of the hallway enables access to any logical information stored in the memory region. The procedure for separating a single logical qubit from the densely packed configuration is given later in Section~\ref{sec:transforming}. Note that the computational region of this microarchitecture is compatible with existing surface code patches layouts (so-called \emph{floorplans}) without dense packing, such as those in Refs.~\cite{beverland2022surface, chamberland2022universal, beverland2022assessing, lee2021even, ueno2024high}.

\section{Transforming Surface Codes into Densely Packed State}
\label{sec:transforming}

In this section, we demonstrate that the dense packing configuration can be achieved through the use of code deformation. The code deformation can be utilized for logical gate~\cite{bombin2009quantum}, but here we perform the deformation only for fusing and splitting codewords.

\begin{figure}[htbp]
    \centering
    \subfloat[][The initial configuration of the surface code patches, which will be densely packed.]{
    \includegraphics[keepaspectratio, width=0.4\linewidth]{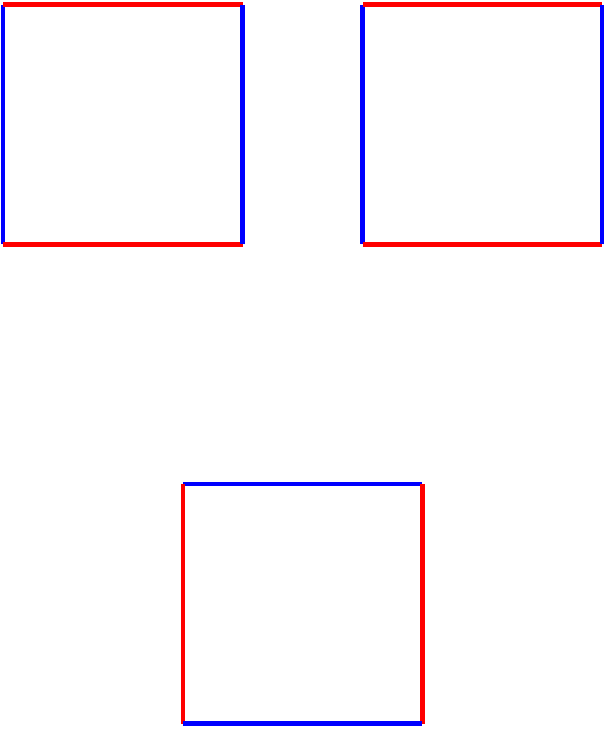}
    }
    \hfill
    \subfloat[][Among the three patches, the top two are deformed. Consequently, the upper patches and the lower single patch face the same boundary. ]{
    \includegraphics[keepaspectratio, width=0.4\linewidth]{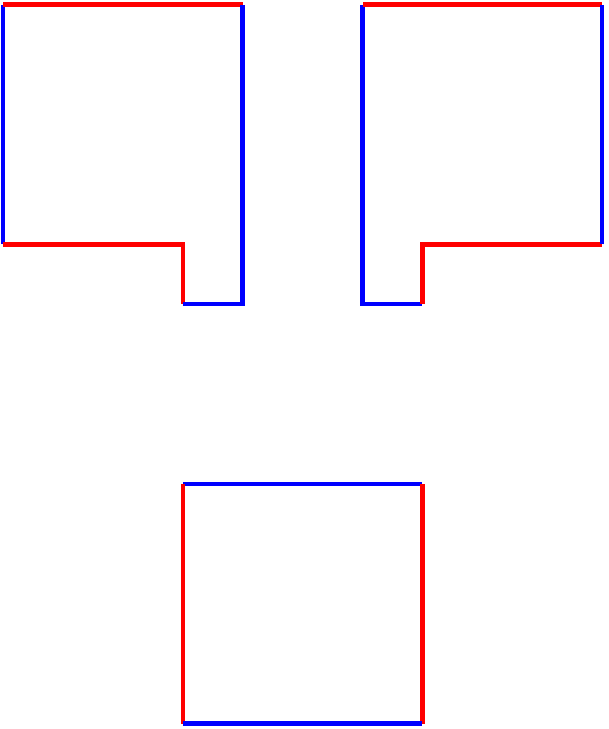}
    }

    \subfloat[][The three codewords have been fused to a single contiguous region, while preserving the original logical states of all three.]{
    \includegraphics[keepaspectratio, width=0.4\linewidth]{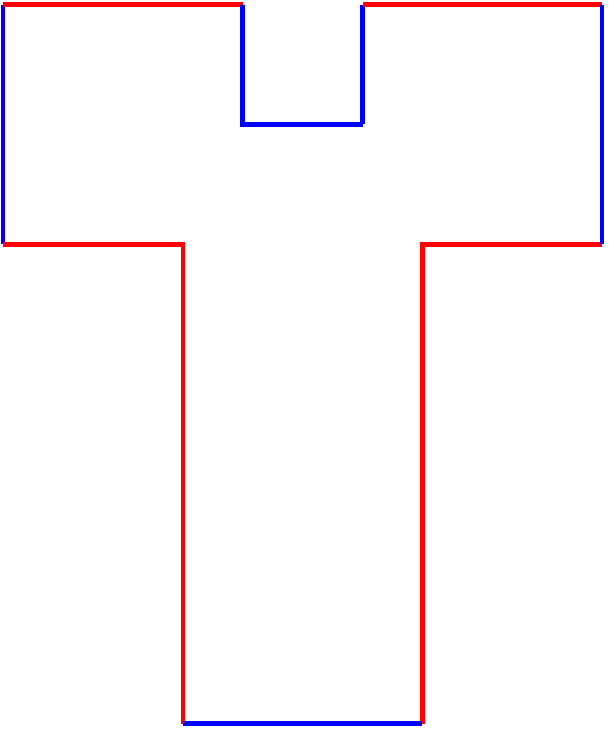}
    }
    \hfill
    \subfloat[][Finally, by reducing the code distance, the dense packing configuration is achieved.]{
    \includegraphics[keepaspectratio, width=0.4\linewidth]{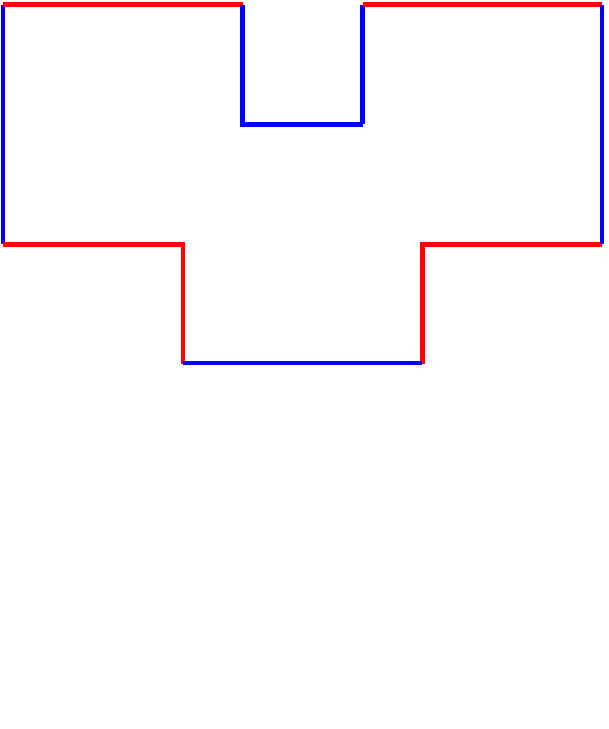}
    }
    \caption{The squares in the figure are schematic representations of surface code patches, illustrating the process by which three surface codes transform into a single dense packing state. The red and blue lines denote the $X$ and $Z$ boundaries, respectively.}
\label{fig:deformation_description_diagram}
\end{figure}

The transformations required for dense packing using code deformation are described below. Fig.~\ref{fig:deformation_description_diagram} illustrates the condensed procedure for deforming the surface code; for a more detailed description of the deformation steps, refer to the Appendix~\ref{sec:app_deform}. In Fig.~\ref{fig:deformation_description_diagram}, the squares represent patches of the surface code, and this figure serves as a simplified picture of Fig.~\ref{fig:bsurface}. The blue lines denote $Z$-boundaries, while the red lines denote $X$-boundaries. The figure schematically illustrates how surface codes are deformed through the use of code deformation.

In Fig.~\ref{fig:deformation_description_diagram}, transforming the patches of the surface code from (a) to (b) and from (b) to (c) each requires $d$ rounds. 
In contrast, the transformation from (c) to (d) can be achieved solely by physical qubit measurement. Therefore, the total time to transform the surface code patches from their standalone configuration to the densely packed configuration is $2d$ rounds plus the time required for physical qubit measurements.
To separate a given patch of the surface code from the densely packed state, the above steps must be performed in the reverse order.  For example, to extract the lower patch of the surface code from (d), the deformation from (d) back to (c) requires $d$ rounds, and the transition from (c) to (b) is performed by physical-qubit measurements. Therefore, the time required to extract a specific surface code from the densely packed state is $d$ rounds plus the time for physical qubit measurements.

\section{Hook-error-avoiding gate scheduling}
\label{sec:gate}

Surface codes require both $X$-type and $Z$-type stabilizer measurements. When performing these measurements, it is necessary to determine appropriate CNOT gate scheduling within the stabilizer measurement circuit.

\begin{figure}[htbp]
    \centering
    \subfloat[][\centering]{
    \includegraphics[keepaspectratio, width=0.35\linewidth]{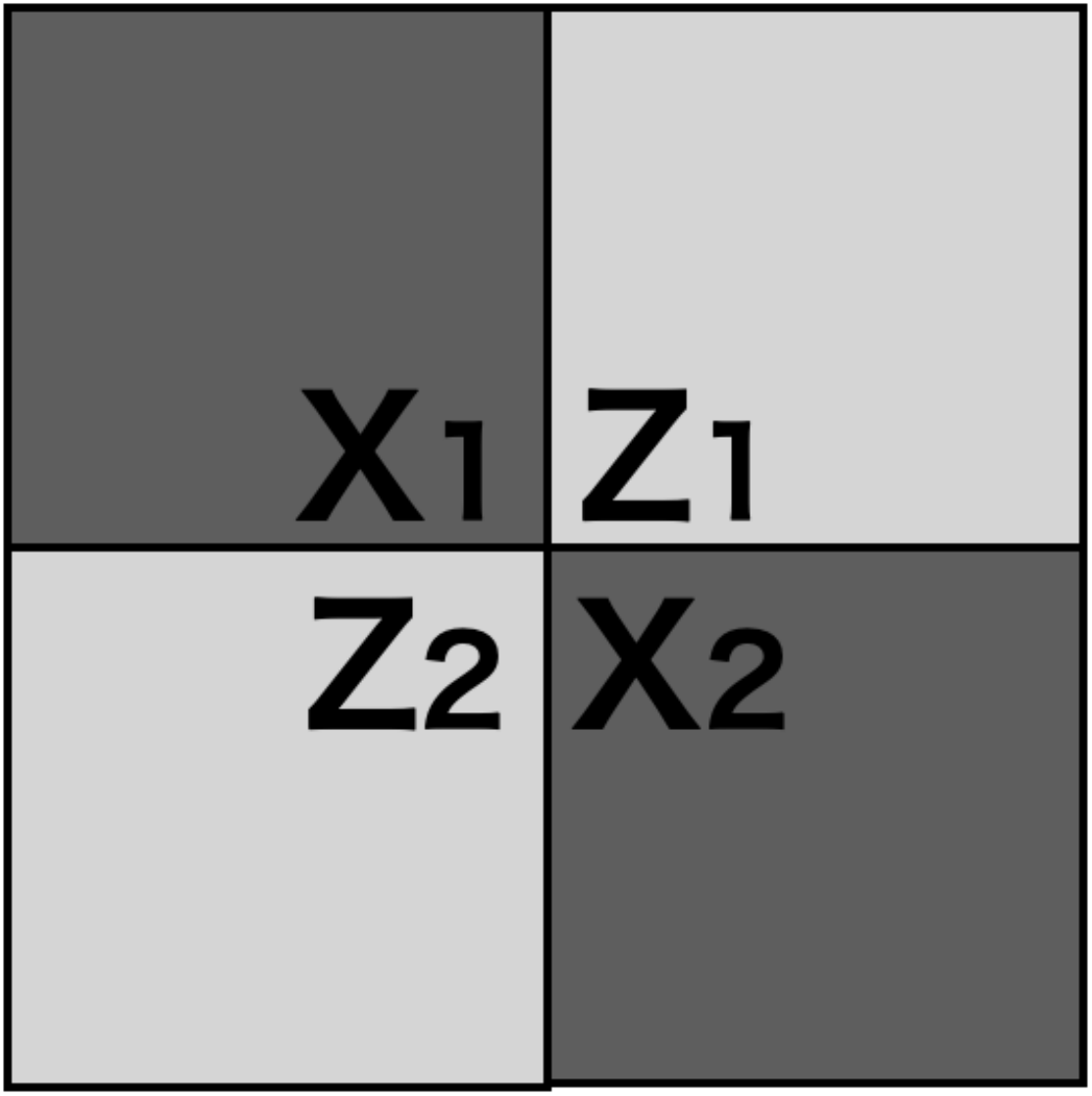}
    }
    \hspace{0.05\textwidth}
    \subfloat[][\centering]{
    \includegraphics[keepaspectratio, width=0.35\linewidth]{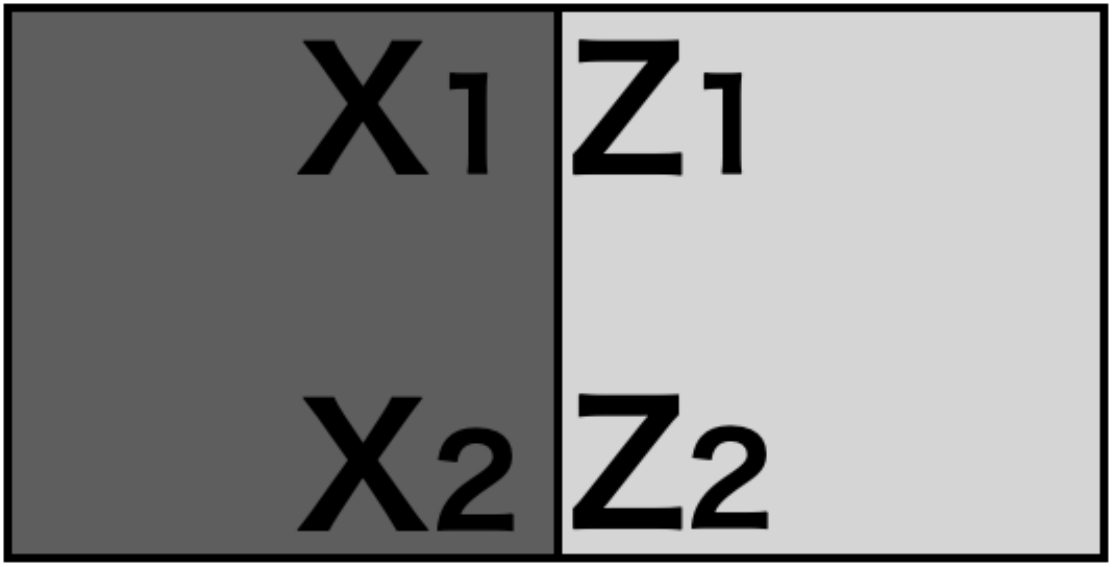}
    }
    \caption{The constraints for gate scheduling in surface codes. (a) Constraint 1. When applying stabilizers on the surface code using CNOT gates, it is not allowed to apply CNOT gates to the same qubit simultaneously. Therefore, the timing of the CNOT gates acting on $X_1$, $X_2$, $Z_1$, and $Z_2$ must not overlap. (b) Constraint 2. The order in which CNOT gates are applied must satisfy either $X_1 < Z_1$ and $X_2 < Z_2$, or $X_1 > Z_1$ and $X_2 > Z_2$.}
\label{fig:gate_schedule12}
\end{figure}

Gate scheduling for stabilizer measurements in the surface code must satisfy two constraints to reproduce correct stabilizer measurements. The first constraint is that when applying gates to data qubits shared by multiple stabilizers, only a single gate can be applied at the same time, as illustrated in Fig.~\ref{fig:gate_schedule12}(a).

The second constraint requires that the sequence of CNOT gate operations for a given stabilizer measurement must not disturb the outcomes of adjacent stabilizer measurements. To satisfy this constraint, it is sufficient to ensure that for the two data qubits shared between neighboring stabilizer measurement circuits, the CNOT gates from one stabilizer measurement circuit always precede those from the other. As illustrated in Fig.~\ref{fig:gate_schedule12}(b), let the time steps at which each stabilizer measurement circuit applies CNOT gates to the data qubits be denoted by $X_1, X_2, Z_1, Z_2$. A valid scheduling requires that either $X_1<Z_1$ and $X_2<Z_2$, or $X_1>Z_1$ and $X_2>Z_2$.

\begin{figure}[htbp]
\includegraphics[width=0.27\textwidth]{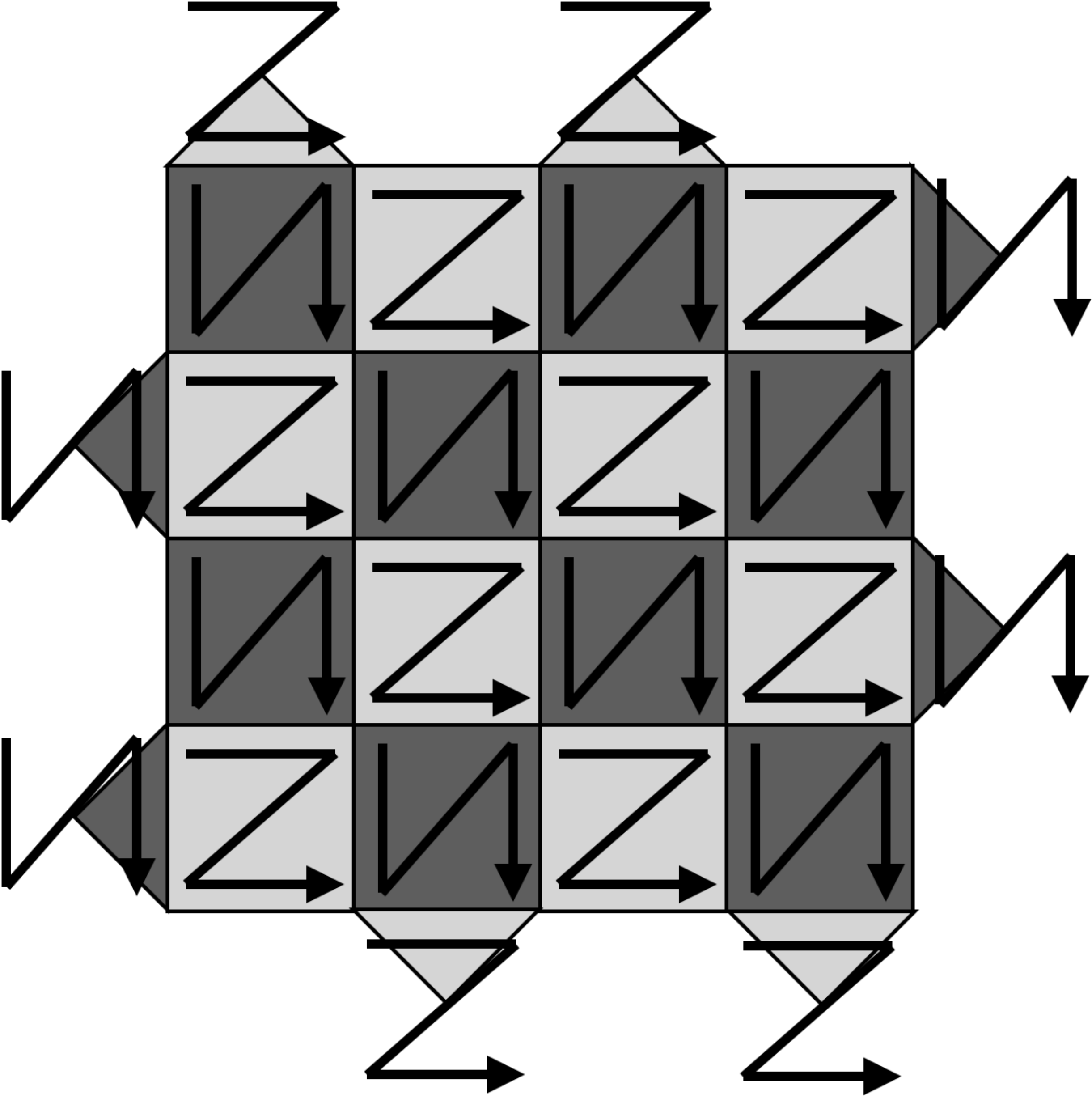}
\caption{\label{fig:normal_gate} Hook-error-avoiding gate scheduling for standalone surface code patch. When applying two‑qubit gates between data qubits at the lattice vertices and the measurement qubits, the gates must be executed in the order indicated by the arrows.}
\end{figure}

Furthermore, considering the suppression of error propagation caused by hook errors in directions parallel to the logical axis, one possible CNOT gate scheduling is as shown in Fig.~\ref{fig:normal_gate}.

When codewords are densely packed, multiple logical qubits are mapped onto a single contiguous surface code patch. As a result, the same-basis logical observables of two connected logical qubits are arranged perpendicularly, necessitating modification of the gate scheduling.

\begin{figure*}[htbp]
\includegraphics[width=0.75\textwidth]{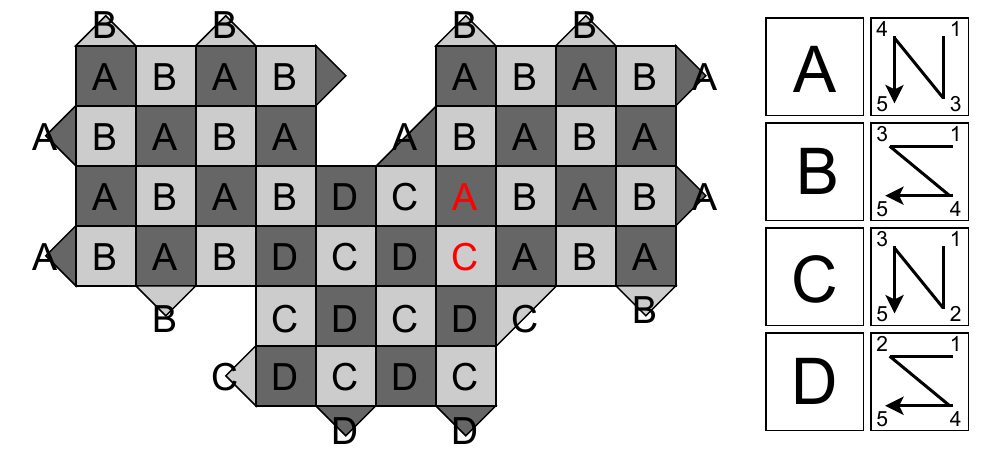}
\caption{\label{fig:gate_scheduling} Gate scheduling for densely packed surface codes. A, B, C, and D represent the order in which the gates are applied. By adjusting the gate scheduling, the propagation of hook errors in the logical direction is suppressed. However, in the area indicated by red text, hook errors still propagate in the logical direction. These stabilizers represent only a small portion of the surface code and are therefore not expected to have a significant impact.}
\end{figure*}

We propose the scheduling shown in Fig.~\ref{fig:gate_scheduling} as a method that fulfills both of the above constraints and largely mitigates the impact of hook errors. In this gate scheduling, hook errors are not entirely prevented from propagating in the logical direction. However, the regions where logical-direction error propagation occurs are limited to a small subset of stabilizers, and their overall impact on the logical error rate is considered to be small.

This gate scheduling consists of five time steps for applying CNOT gates, resulting in a longer stabilizer measurement duration compared to standalone surface code gate scheduling. While this might suggest an increase in the error rate compared to the standalone surface code, in practice, the error rate remains effectively suppressed even in a densely packed configuration. In Section~\ref{sec:simulation}, we verify this through simulations employing this gate scheduling in a densely packed arrangement.

\section{Error rates simulation of densely packed surface code}
\label{sec:simulation}

To estimate the circuit-level logical error rate of the densely packed surface codes, we constructed quantum circuits using the Python library stim~\cite{gidney2021stim}, and performed syndrome decoding using pymatching~\cite{higgott2023sparse}. The codes used for the simulation are available in Ref.~\cite{mycode}.

\begin{figure}[htbp]
\includegraphics[width=0.4\textwidth]{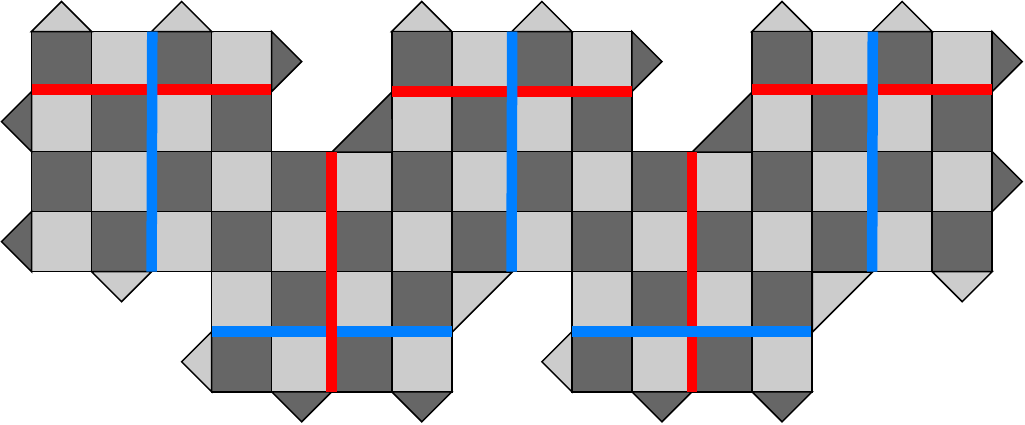}
\caption{\label{fig:simulation_dense} The densely packed patch encoding $5$-logical qubits that is used in the error correction simulation. The simulation focused on evaluating the error rate of only the central logical qubit. The red and blue lines correspond to the logical $X$ and $Z$ operators of the logical qubits, respectively.}
\end{figure}

We performed three types of simulations. The first simulation considered the densely packed surface code patches that adopt the gate scheduling shown in Fig.~\ref{fig:gate_scheduling}. As illustrated in Fig.~\ref{fig:simulation_dense}, we constructed a densely packed surface code of five codewords and evaluated the logical error rates of all five logical qubits. 
We refer to this densely packed surface code as hook-avoiding dense.
At the end of the error correction procedure, all physical qubits were destructively measured to verify whether logical errors occurred along the logical operators indicated by the red and blue lines.

The second simulation also examined the densely packed state, but employing a gate scheduling that does not account for hook errors. 
We refer to this densely packed surface code as hook-prone dense.
Same as the first case, a five-patch densely packed surface code was constructed as shown in Fig.~\ref{fig:simulation_dense}. In this surface code, the scheduling of the $X$-type stabilizers was fixed to pattern B in Fig.~\ref{fig:gate_scheduling}, while the scheduling of the $Z$-type stabilizers was fixed to pattern A. Both schedules (A and B) were adapted to four time steps. Under this scheduling, the two codewords at the bottom of Fig.~\ref{fig:simulation_dense} are protected from hook errors, whereas the three codewords at the top remain vulnerable to them.

\begin{figure*}[htbp]
    \centering
    \subfloat[][Logical $X$ error rate per round of central logical qubit.]{
    \includegraphics[keepaspectratio, width=0.47\linewidth]{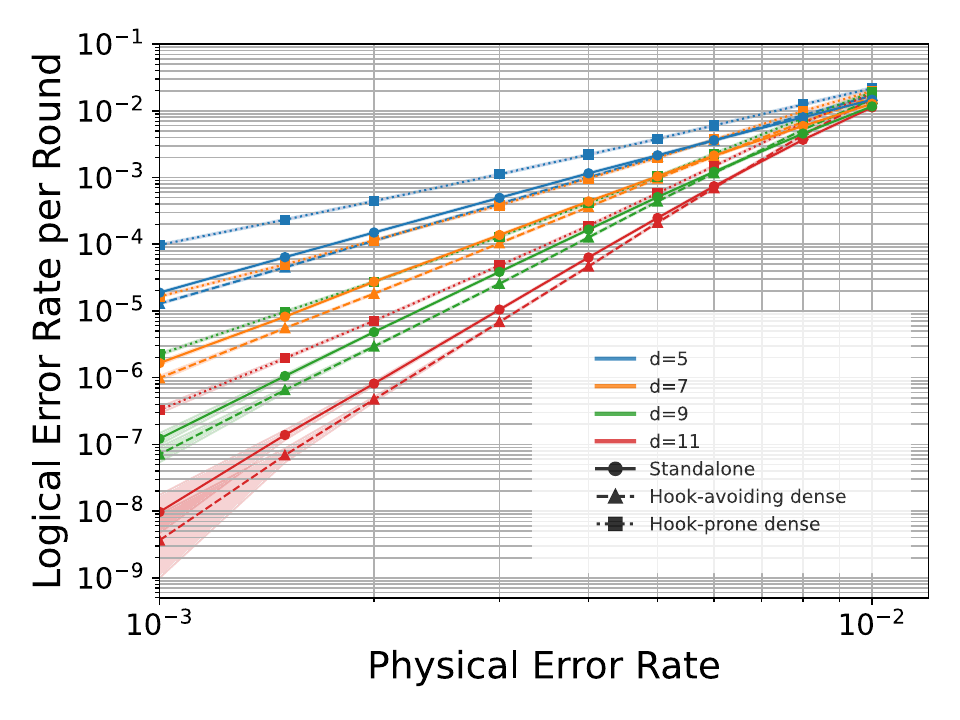}
    }
    \hfill
    \subfloat[][Logical $Z$ error rate per round of central logical qubit.]{
    \includegraphics[keepaspectratio, width=0.47\linewidth]{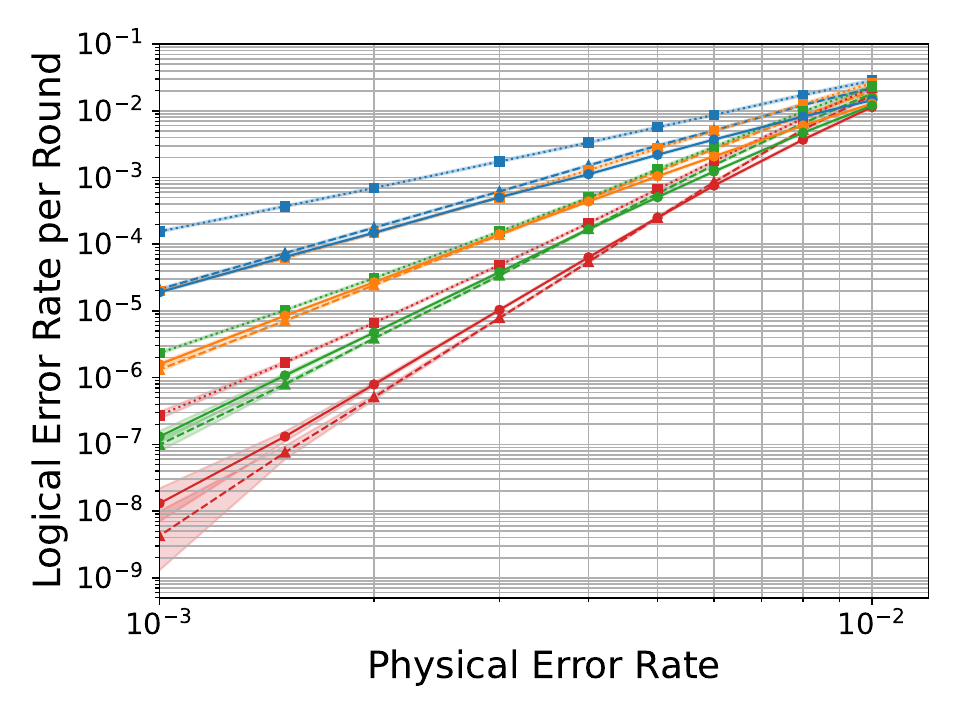}
    }\\
    \subfloat[][Logical $X$ error rate per round of second from the left logical qubit.]{
    \includegraphics[keepaspectratio, width=0.47\linewidth]{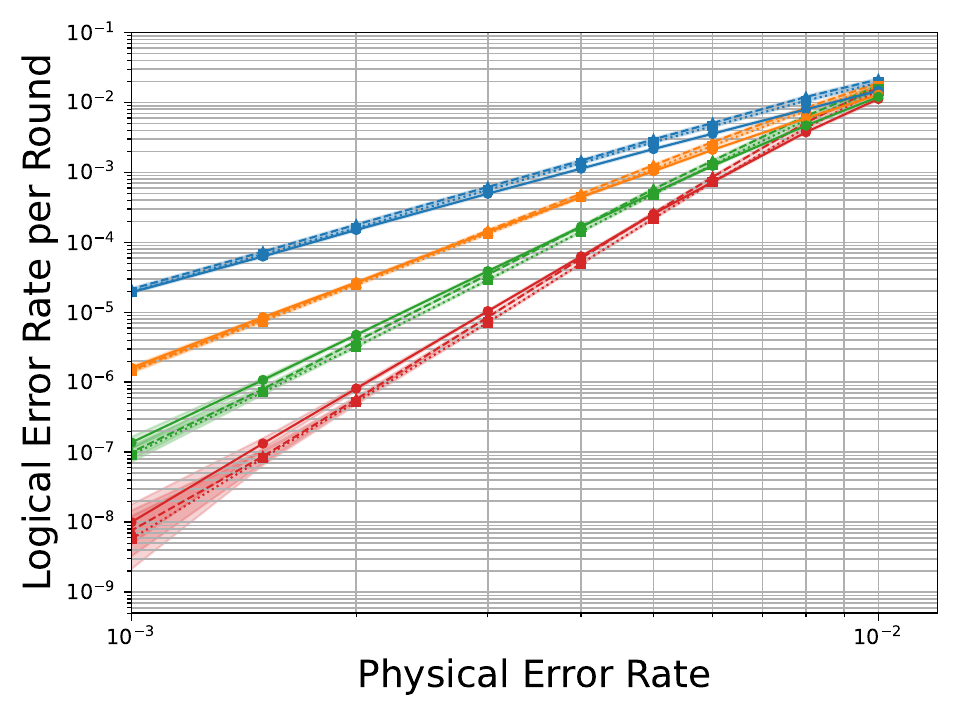}
    }
    \hfill
    \subfloat[][Logical $Z$ error rate per round of second from the left logical qubit.]{
    \includegraphics[keepaspectratio, width=0.47\linewidth]{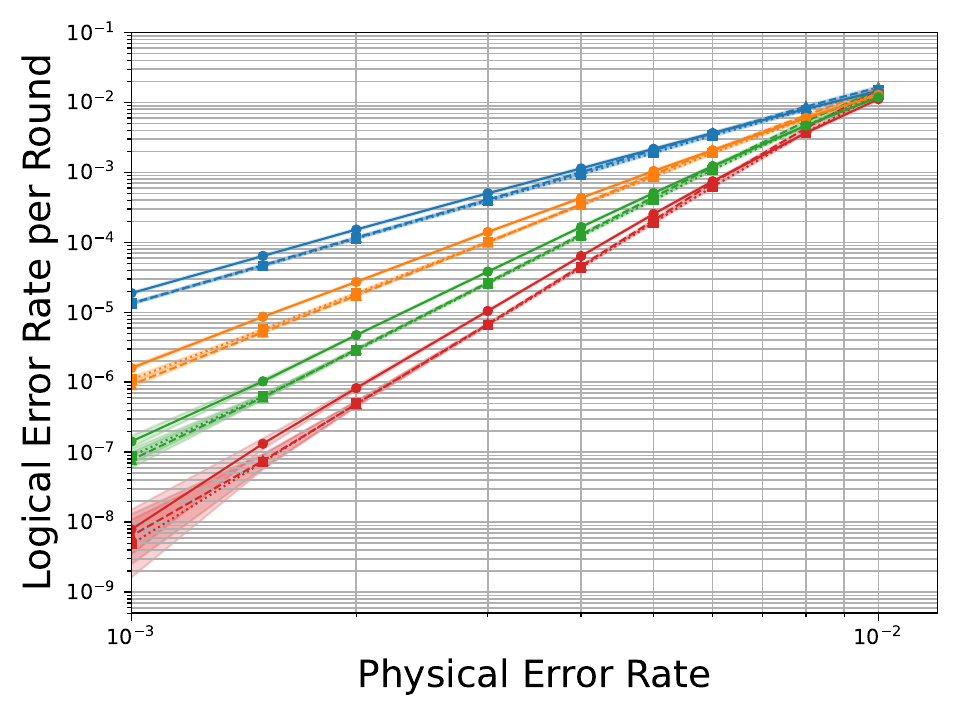}
    }
    \caption{Simulation results of logical error rates of surface codes with different configurations. The horizontal axis shows the physical qubit error rate, and the vertical axis shows the logical error rate per round. Error correction was performed for $3d$ rounds. The solid, dashed, and dotted lines correspond to the standalone surface code, hook-avoiding dense, and hook-prone dense, respectively. Hook-avoiding dense adopts the gate scheduling in Fig.~\ref{fig:gate_scheduling}, while hook-prone dense uses a scheduling that does not account for hook errors. The lightly shaded region indicates where the Bayes factor (likelihood ratio) reaches $1000$.}
\label{fig:simulation_result}
\end{figure*}

The third simulation evaluated the logical error rates of standalone surface code, not densely packed. In this case, the scheduling of the $X$-type stabilizers was fixed to pattern A, and that of $Z$-type stabilizers was fixed to pattern B, with both schedulings adapted to four time steps. These schedulings correspond to a conventional scheme designed to suppress the impact of hook errors.

This is achieved by simulating the logical error rate corresponding to $X$ errors on the logical $\ket{0}$ state and $Z$ errors on the logical $\ket{+}$ state. 
We evaluated both the logical $X$ error rate and logical $Z$ error rate because there exists asymmetry in the susceptibility to logical X errors and Z errors for each codeword on the densely packed surface code. Focusing on the orientation of the logical $Z/X$ operators of the central codeword in Fig.~\ref{fig:simulation_dense}, we can see that graphical asymmetry can lead to asymmetry between the logical $X$ error rate and the logical $Z$ error rate. When extending the densely packed surface code in Fig.~\ref{fig:simulation_dense} further, the codewords in the upper region will exhibit the same asymmetry as the central codeword, while the codewords in the bottom region will exhibit reversed $X/Z$ error rate asymmetry.

The maximum number of shots is set to $10^8$, and the maximum number of errors is limited to $10^4$.

The noise model used in the simulation assumes that two-qubit gates, measurement, and initialization operations are the noisy processes. The details are as follows:
\begin{enumerate}
\item Each CNOT gate is followed by two-qubit depolarizing noise with a probability $p$.
\item Each measurement operation is followed by a result flip with a probability $p$ and single-qubit depolarizing noise with a probability $p$.
\item Each initialization operation is followed by a result flip with a probability $p$.
\item Each single-qubit gate is followed by single-qubit depolarizing noise with a probability $p/10$.
\item Each idle qubit location(where no gate operation occurs) is followed by single-qubit depolarizing noise with a probability $p/10$ during each operation.
\end{enumerate}

Single-qubit depolarizing noise with a probability $p$ means that during each single-qubit gate operation, an $X,Y,Z$ error occurs on the qubit with a probability of $p/3$ each, while no error occurs with a probability of $1-p$. Two-qubit depolarizing noise with a probability $p$ means that during each two-qubit gate operation, one of the \{$I\otimes X, I\otimes Y, I\otimes Z, X\otimes I, X\otimes X, X\otimes Y, X\otimes Z, Y\otimes I, Y\otimes X, Y\otimes Y, Y\otimes Z, Z\otimes I, Z\otimes X, Z\otimes Y, Z\otimes Z$\} errors occurs with a probability of $p/15$ each, while no error occurs with a probability of $1-p$.

The simulation results are shown in Fig.~\ref{fig:simulation_result}. Although we evaluated logical error rates of all logical qubits, we present simulation results only for the central and the second-from-the-left logical qubits. This is because the logical qubits can be classified into two groups: leftmost, central, and rightmost logical qubits exhibit the same scaling behavior of the logical error rate, while the second-from-the-left and second-from-the-right logical qubits share another same scaling behavior. 

We first examine the logical error rate of the central logical qubit. We compare the logical error rates of two types of densely packed surface codes. 
We found that hook-avoiding dense yielded lower logical error rates than hook-prone dense.
Then, we compare the logical error rates of the hook-avoiding dense and the standalone surface code. We observed that the standalone surface code patches generally yielded lower logical error rates when the physical error rate is around $10^{-2}$. However, as the physical error rate decreases, the hook-avoiding dense exhibited lower logical error rates than the standalone surface code.

In the hook-avoiding dense, the number of times slices of CNOT gates increases from four to five compared to the standalone. This increases the likelihood of physical errors on idle qubits. However, for code distances $d\geq 7$, the logical error rates of the hook-avoiding dense become lower than that of the standalone patch as the physical error rate decreases. This improvement can be attributed to the fact that, in the hook-avoiding dense configuration, multiple patches share common regions where logical errors are less likely to occur. In standalone surface code patches, a logical error can occur with as few as $(d+1)/2$ physical errors. In contrast, in the hook-avoiding dense, a logical error that spans across the fused regions requires more than $(d+1)/2$ physical errors. Since the regions where logical errors can occur by only $(d+1)/2$ physical errors are limited to non-fused parts. 

On the other hand, the higher logical error rate observed in the hook-prone dense is likely due to the increased susceptibility to hook errors. The number of times CNOT gates are applied for stabilizer measurements in this gate scheduling is the same as that in the standalone patch, and, same as hook-avoiding dense, multiple patches share common regions. However, the logical error rates increase because hook errors propagate in the same direction as the logical operators.

Next, we consider the logical error rate of the second-from-the-left logical qubit. For this qubit, the relationship between the logical error rates of the standalone surface code and the hook-avoiding dense configuration is consistent with that observed for the central qubit. In contrast, we found that the logical error rate of the hook-prone dense configuration exhibits behavior very similar to that of the hook-avoiding dense configuration. This behavior can be attributed to the orientation of the logical 
$X/Z$ operators of the second-from-the-left qubit, which are rotated by 90 degrees relative to those of the central qubit. As a result, hook errors no longer propagate parallel to the logical operators, suppressing their contribution to logical failures.

However, in cases where hook errors influence the logical error rate, such as for the central logical qubit, the degradation of the logical error rate in the hook-prone dense configuration becomes pronounced. Therefore, when considering the overall logical error performance of the densely packed state, it is preferable to adopt the hook-avoiding dense gate scheduling. 

Finally, in a densely packed configuration, correlated logical errors between multiple logical qubits may occur. However, since the logical error rates of all logical qubits in the hook-avoiding dense configuration are comparable to or lower than those of the standalone surface code, we expect that any such correlations do not significantly degrade the overall logical error performance.

In this work, we present the logical error rates of the central and second-from-the-left logical qubits. The logical error rates of all logical qubits in the densely packed configuration can be evaluated using the code provided in Ref.~\cite{mycode}.

This simulation evaluated the logical error rates only for code distances of 5, 7, 9, and 11. However, based on these results, it is expected that the hook-avoiding dense surface code will exhibit logical error rates comparable to or lower than the standalone surface code, even with larger code distances.

\FloatBarrier

\section{Conclusion}
\label{sec:conclusion}
 
In this paper, we propose an implementation method for dense packing of surface code patches.
It was confirmed that the area required for densely packed surface code codewords asymptotically becomes three-fourths of that of standalone patches. We demonstrated that dense packing from standalone patches can be achieved through code deformation with $2d$ rounds plus the time required for physical qubits measurements, while extracting a logical qubit requires $d$ rounds plus the time for physical qubits measurements.

The surface code in a densely packed state employs a gate scheduling different from the standard surface code. To minimize error propagation along the logical operator direction caused by hook errors, we propose to employ four types of gate scheduling for each stabilizer generator. In this approach, there exists one point per logical qubit where a hook error can propagate parallel along the logical operator direction. 
However, the region of the densely packed surface code affected by this error propagation is limited, and it is expected to have little impact on the logical error rate.
This observation was verified through simulations, showing that even in a densely packed state, increasing the code distance and reducing the physical error rate result in a logical error rate that is comparable to or lower than that of the standalone surface code. 
Multiple logical qubits share part of the patch regions that would otherwise be independently assigned to each of them. In a standalone surface code, a logical error can occur with $(d+1)/2$ physical qubit errors. However, since densely packed surface code shares a patch region, the minimum number of physical qubit errors required to induce a logical error in that part increases. We believe this fact contributes to reducing the logical error rate of densely packed surface code. As the code distance increases and the physical error rate decreases, the logical error rate of densely packed surface code falls below the logical error rate of standalone surface code, and we believe this is due to the influence of this factor.

A practically important direction is the evaluation of the microarchitecture with a hierarchical memory approach. 
Our microarchitecture has the potential to reduce the spatio-temporal overhead of fault-tolerant quantum computing by employing space-efficient dense packing in the memory region and time-efficient standalone surface codes in the computational region. Evaluating the performance and its compatibility with various surface-code computing layouts~\cite{litinski2019game, kobori2025lsqca, hirano2025locality} remains an important topic for future work.

\section*{Acknowledgment}
This work was supported by
JST Moonshot R\&D Grant Number JPMJMS226C 
and the New Energy and Industrial Technology Development Organization (NEDO) JPNP23003. 
Shin Nishio has been supported by JSPS Overseas Research Fellowships.

\newpage
\appendix

\section{Surface Code Deformations for the dense packing at the Physical Qubit Level}
\label{sec:app_deform}
We describe the exact deformation procedures from the standard surface code patches to the densely packed state at the physical qubit level. These deformations correspond to the condensed procedure shown in Fig.~\ref{fig:deformation_description_diagram}.

First, we prepare multiple standalone surface code patches and physical $\ket{0}$ states as shown in Fig.~\ref{fig:deformation_description_physical}(a). Without applying any physical operation, new stabilizer generators adjacent to the two upper patches can be defined as shown in Fig.~\ref{fig:deformation_description_physical}(b). The eigenvalues of these new stabilizer generators are already determined because the newly added qubits are in $\ket{0}$ states. Then, we perform $X$ and $Z$ stabilizer measurements on the patches and the new stabilizer generators depicted in Fig.~\ref{fig:deformation_description_physical}(c). 
After $d$ rounds of this type of stabilizer measurement, the three patches stabilize in the state shown in Fig.~\ref{fig:deformation_description_physical}(c). As a result, the Z boundaries of the upper two patches face the Z boundary of the lower patch. 
Simultaneously with the completion of the deformations of the upper two patches, we initialize the physical qubits in $\ket{+}$ states as shown in Fig.~\ref{fig:deformation_description_physical}(d).
Without performing any physical operations, new $X$-type stabilizer generators can be defined with the physical qubits prepared in the $\ket{+}$ state as their vertices. The newly defined stabilizer generators are shown in  Fig.~\ref{fig:deformation_description_physical}(e). 
Then, we perform $X$ and $Z$ stabilizer measurements, including the new stabilizer generators depicted in Fig.~\ref{fig:deformation_description_physical}(f). 
After $d$ rounds of this type of stabilizer measurements, the three patches fuse together as shown in Fig.~\ref{fig:deformation_description_physical}(f). Finally, we measure the data qubits of the lower long portion of the fused patches to shorten its length. 
Eventually, three patches are fused into the densely packed configuration as shown in Fig.~\ref{fig:deformation_description_physical}(g).

To separate patches from a densely packed state, one can achieve this by reversing the above procedure. 
In doing so, the operations to initialize the physical qubits to the $\ket{0}$/$\ket{+}$ must be replaced with the operations to measure the physical qubits in the $Z$/$X$ basis. 

\begin{figure*}[b]
    \centering
    \subfloat[][The initial states for performing dense packing. The black dots represent physical qubits. These three standalone surface code patches subsequently undergo deformations to achieve the denser configuration. ]{
    \includegraphics[keepaspectratio, width=0.45\linewidth]{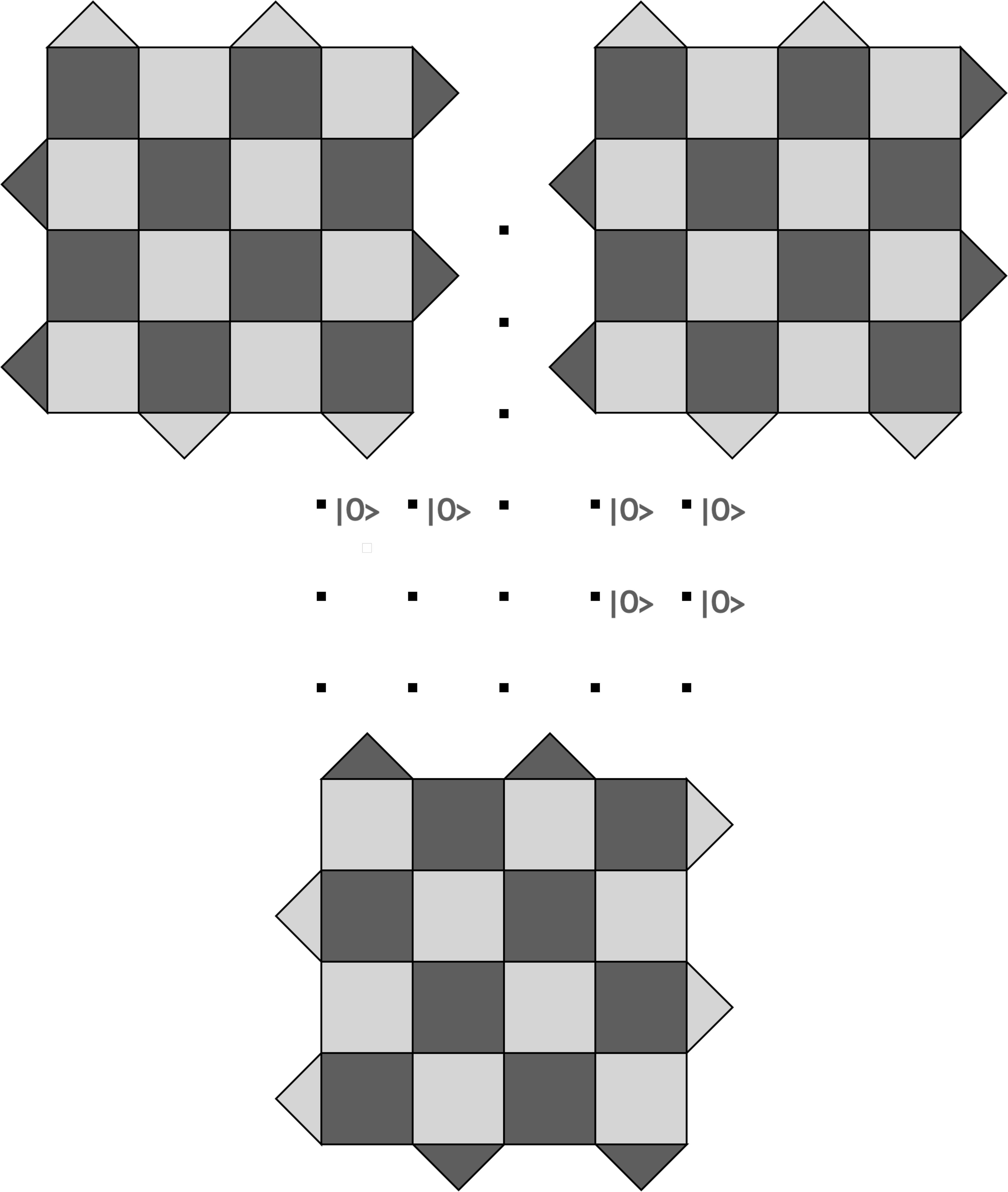}
    }
    \hfill
    \subfloat[][The stabilizer generators associated with new data qubits which are in $\ket{0}$ state in (a) are determined without performing any measurements.]{
    \includegraphics[keepaspectratio, width=0.45\linewidth]{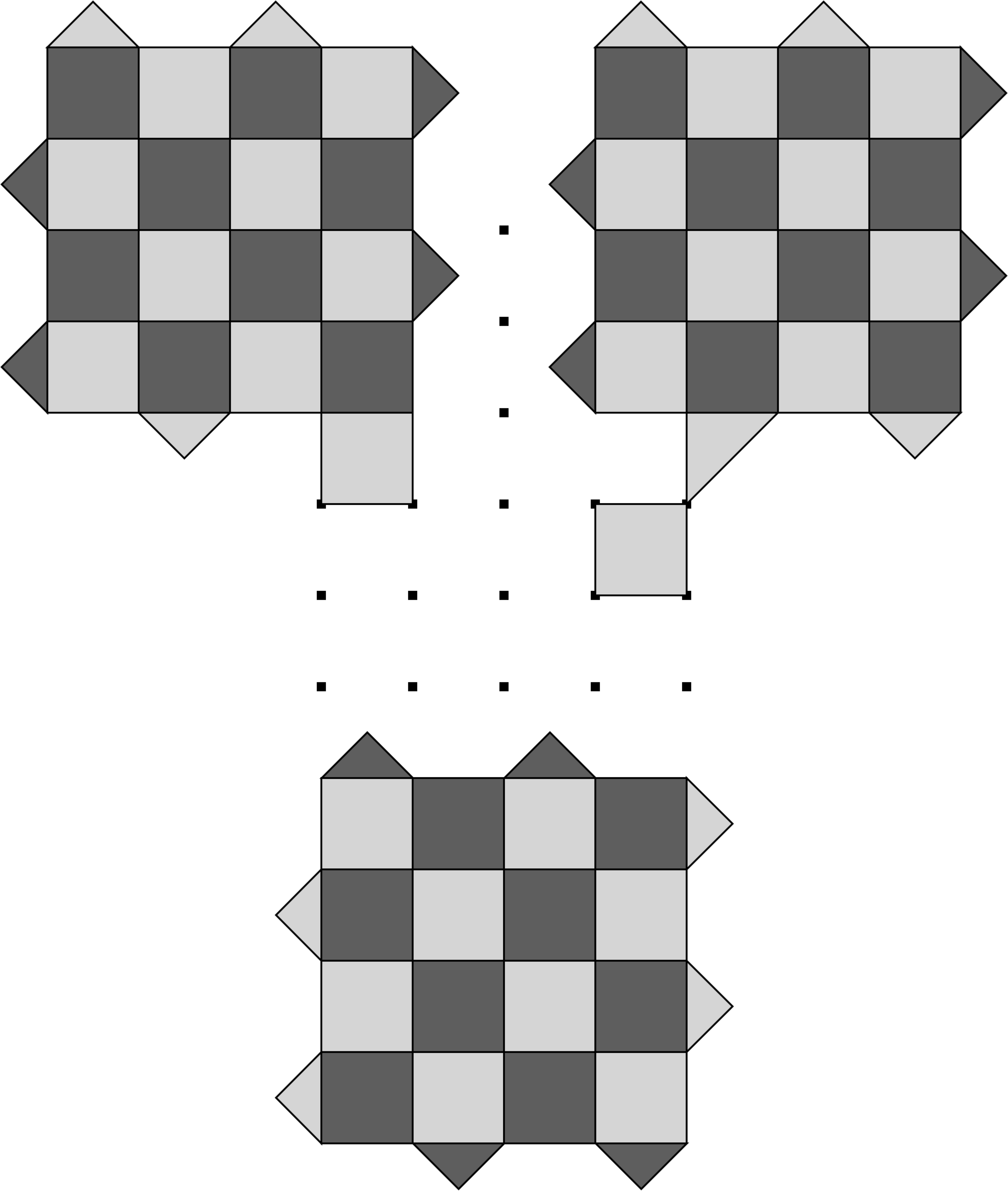}
    }
    \hfill
    \subfloat[][The upper two patches have been deformed, and the $Z$ boundaries of these two patches are adjacent to the $Z$ boundary of the lower patch.]{
    \includegraphics[keepaspectratio, width=0.45\linewidth]{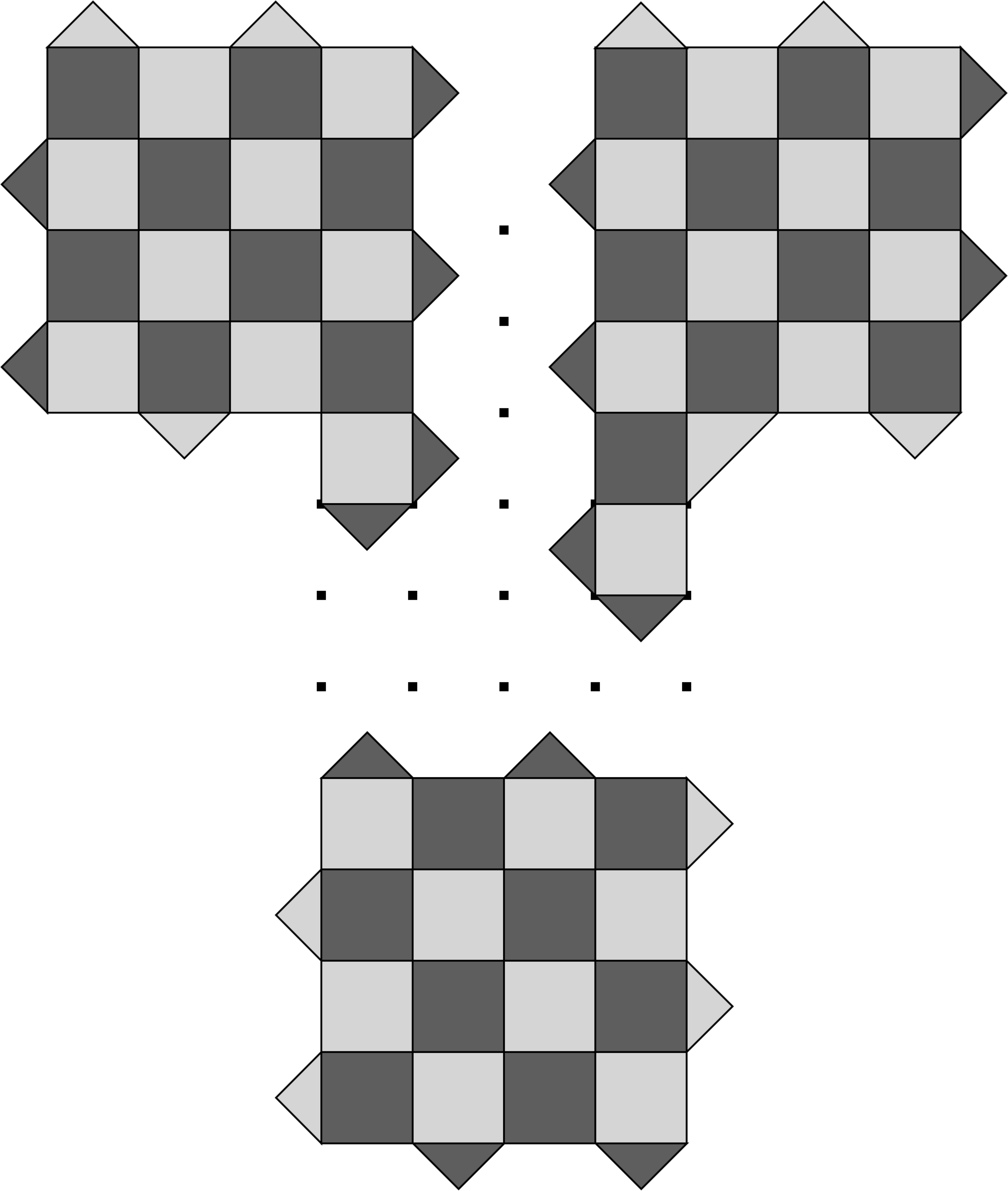}
    }
    \hfill
    \subfloat[][Two deformed surface code patches and a standalone surface code patch. Physical qubits are in $\ket{+}$ states.]{
    \includegraphics[keepaspectratio, width=0.45\linewidth]{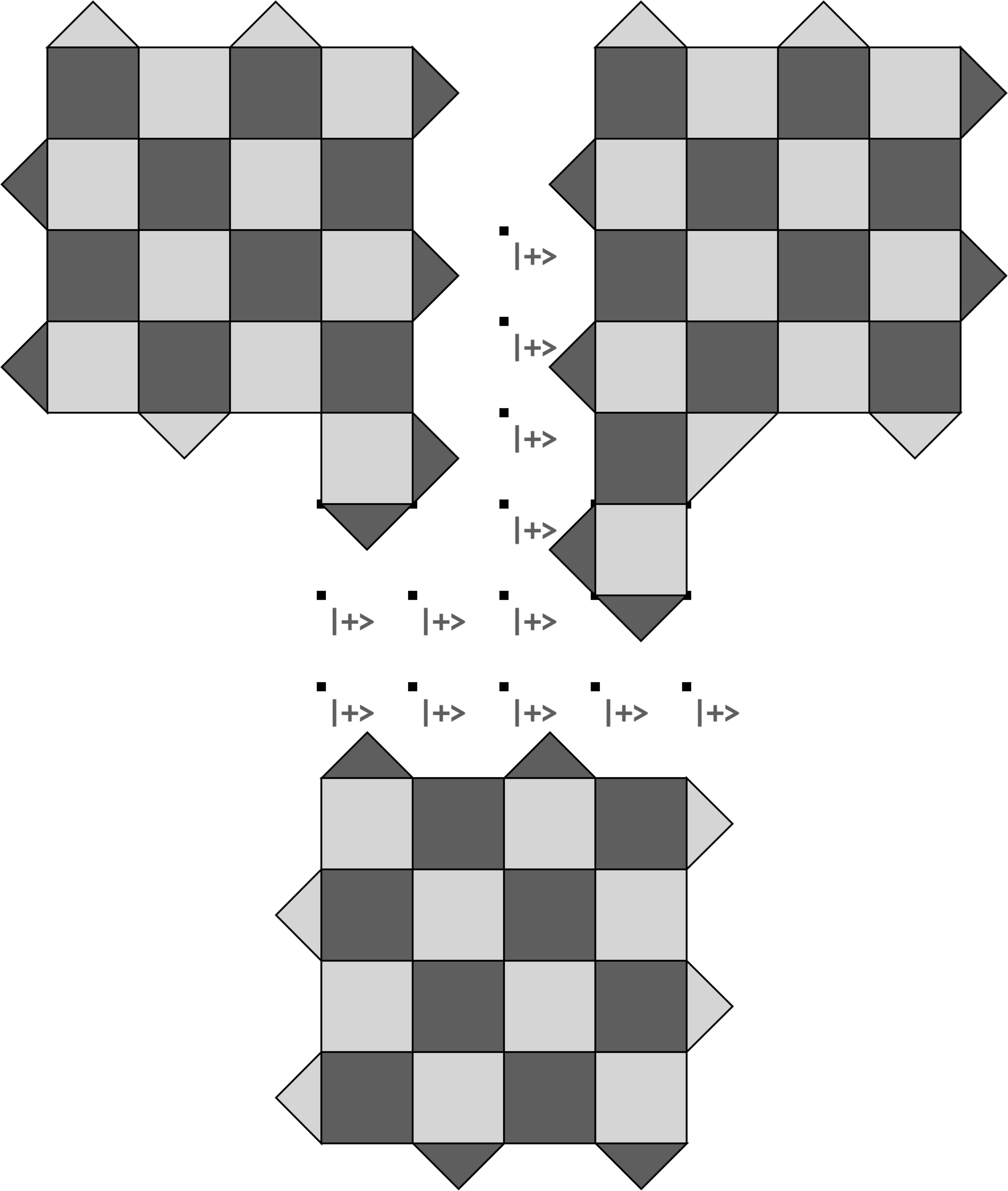}
    }
    \caption{Code deformation procedure for dense packing of the surface code}
\label{fig:deformation_description_physical}
\end{figure*}

\begin{figure*}
\ContinuedFloat
    \raggedright
    \subfloat[][The newly added dark shaded squares represent $X$-type stabilizers, which can be defined without stabilizer measurements.]{
    \includegraphics[keepaspectratio, width=0.45\linewidth]{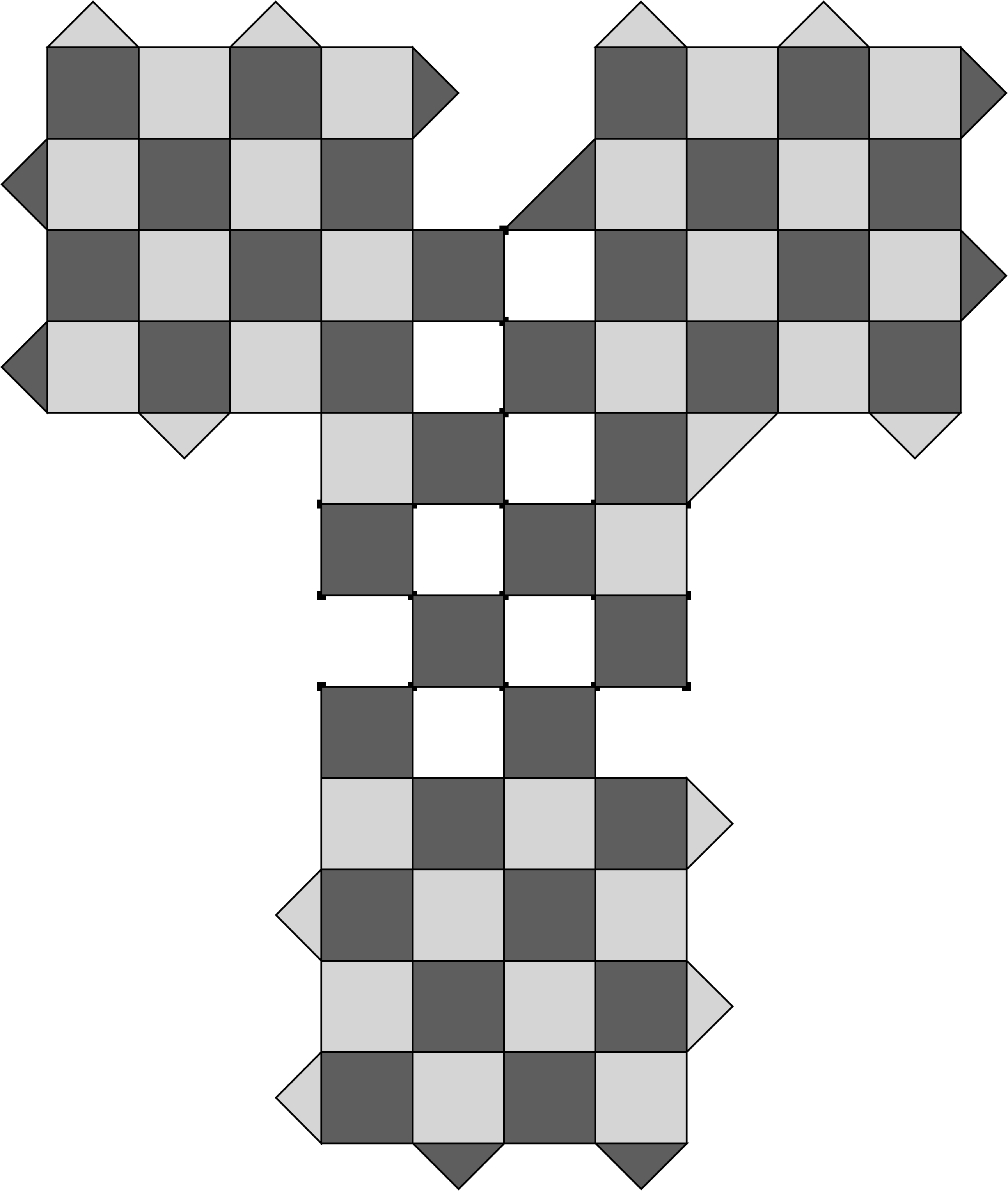}
    }
    \hfill
    \subfloat[][The state where three patches are fused together.]{
    \includegraphics[keepaspectratio, width=0.45\linewidth]{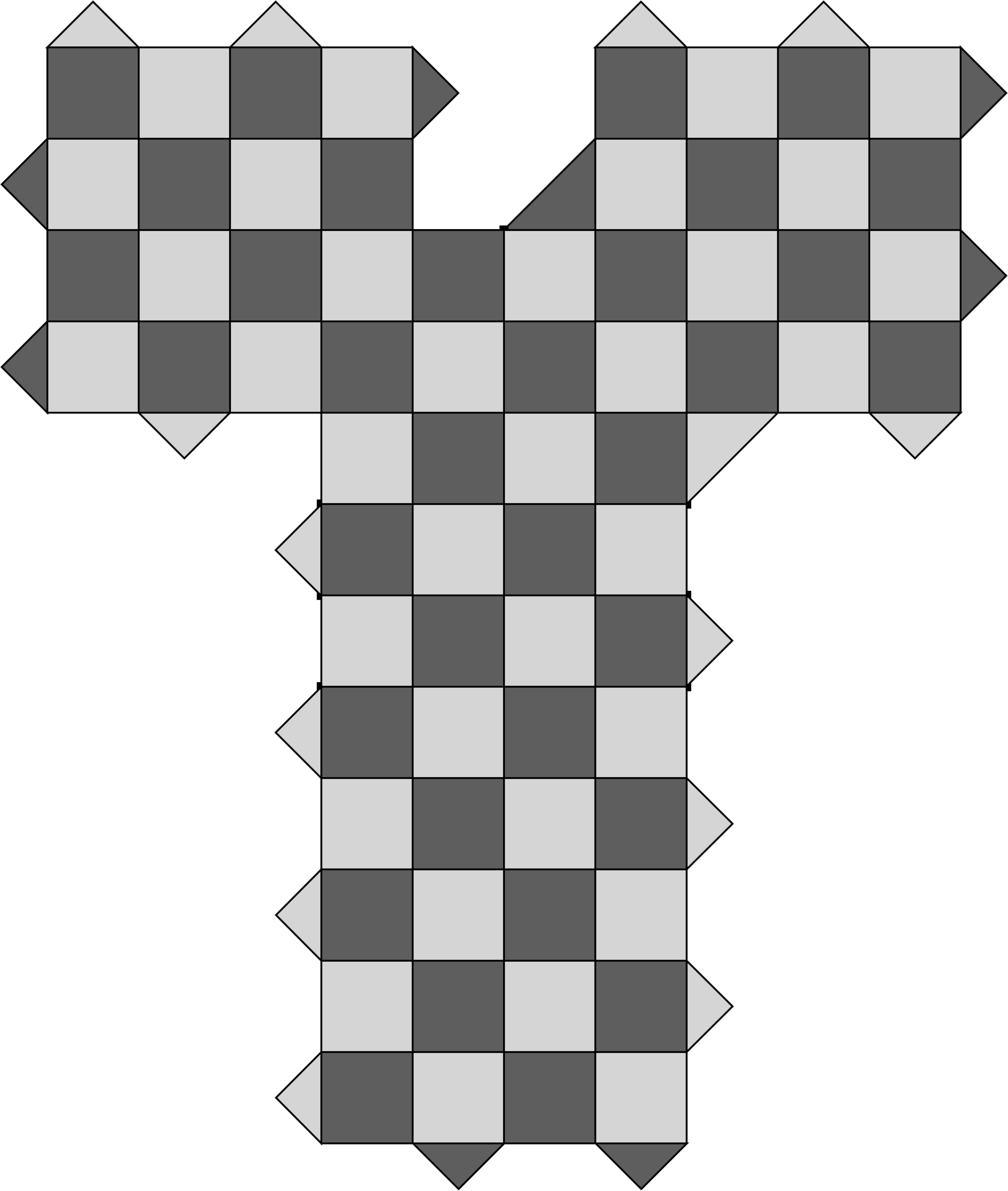}
    }

    \subfloat[][The lower part has become shorter from the (f) state and the three patches are in a densely packed state.]{
    \includegraphics[keepaspectratio, width=0.45\linewidth]{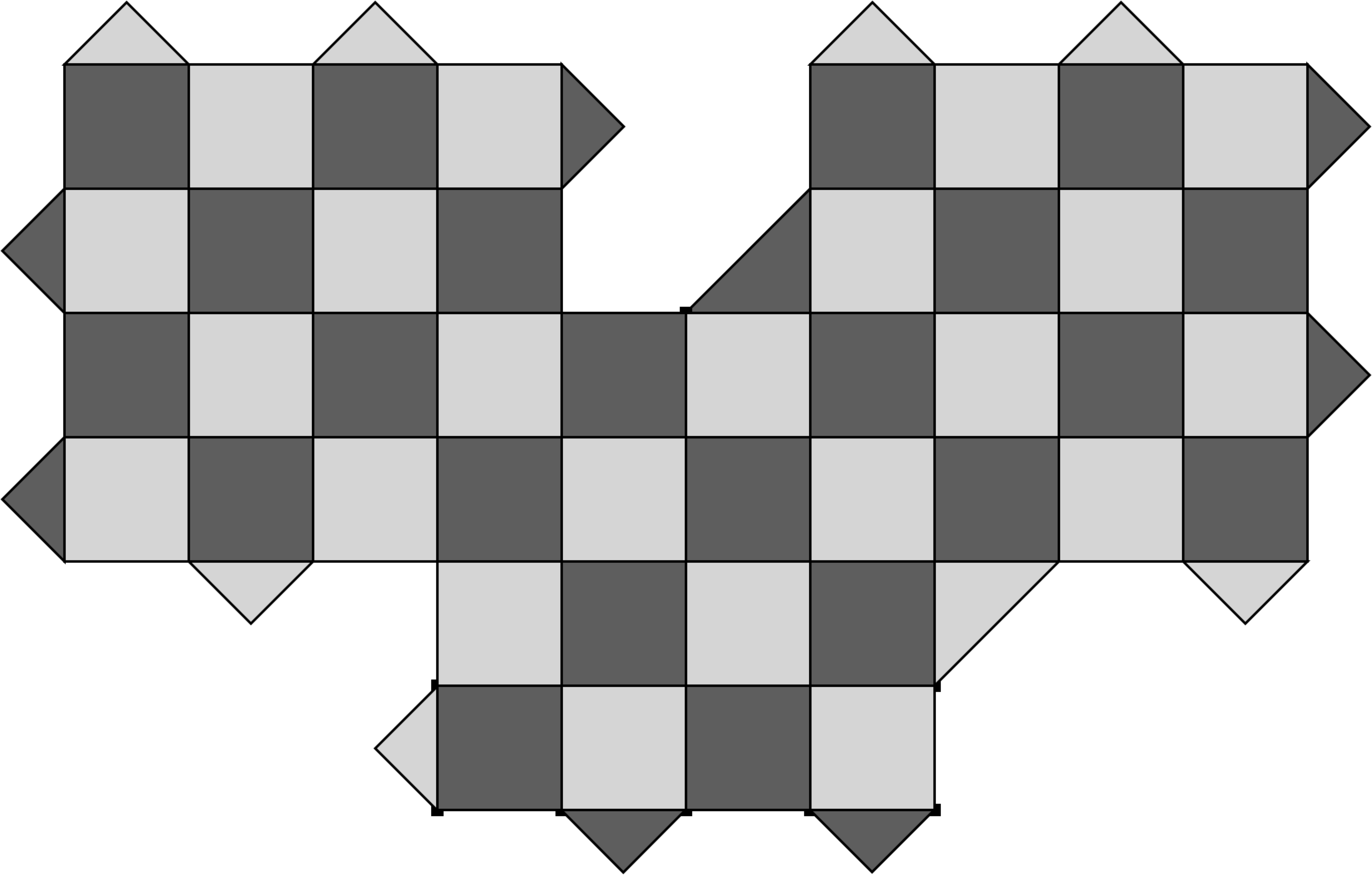}
    }
    
    \caption{Code deformation procedure for dense packing of the surface code}
\label{fig:deformation_description_physical}
\end{figure*}
\clearpage
\section{Movement method for densely packed surface code}
\label{sec:app_movement}

\begin{figure*}[htbp]
     \raggedright
    \subfloat[][Densely packed surface code encoding three logical qubits. The space below the codes is the physical qubits used as a hallway.]{\raisebox{9.7em}{
    \includegraphics[keepaspectratio, width=0.3\linewidth]{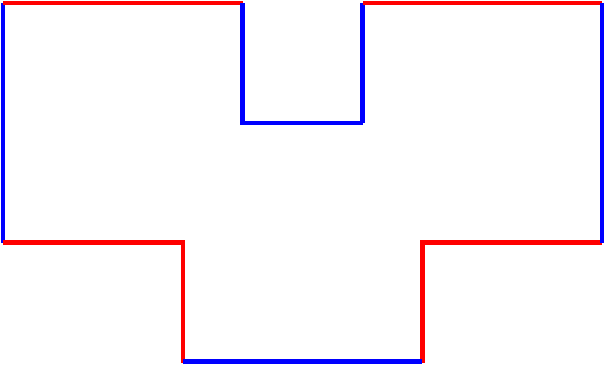}
    }}
    \hspace{1em}
    \subfloat[][The lower part of the densely packed surface code is extended downward.]{
    \includegraphics[keepaspectratio, width=0.3\linewidth]{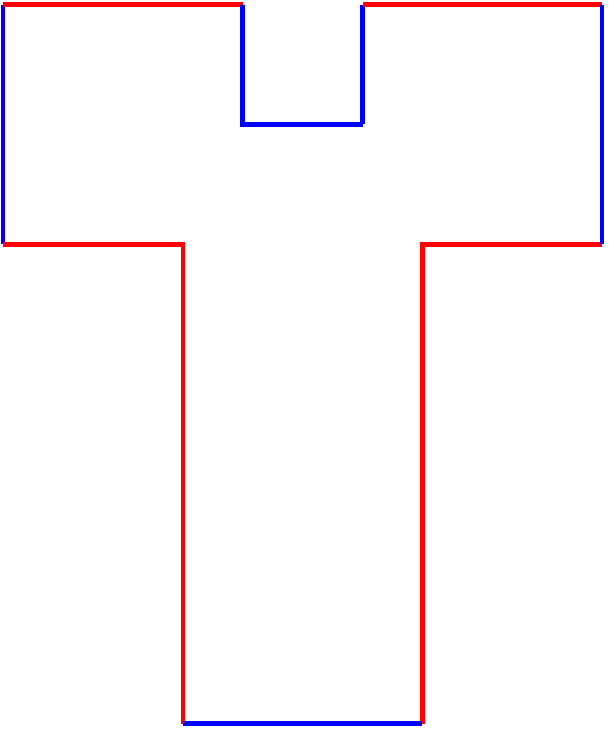}
    }
    \hspace{1em}
    \subfloat[][The upper region is expanded downward.]{
    \includegraphics[keepaspectratio, width=0.3\linewidth]{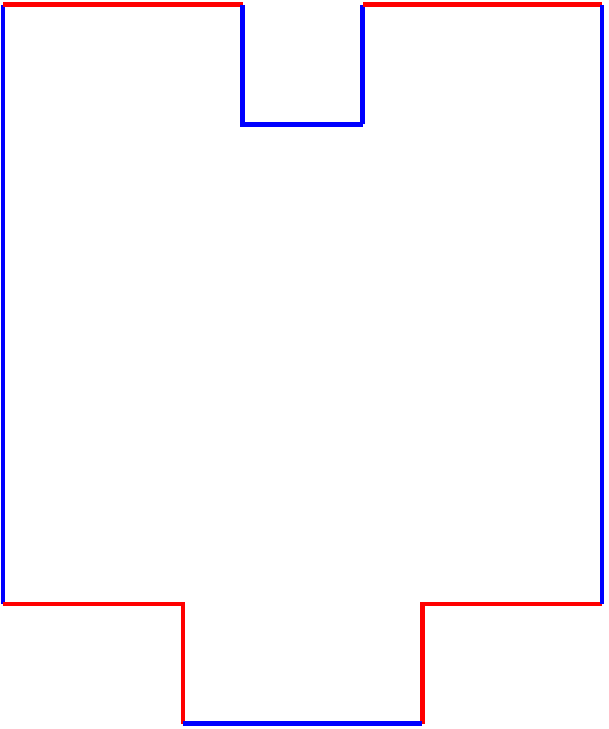}
    }
    \\[10mm]
    \subfloat[][The upper $Z$ boundary has moved closer to the lower region.]{
    \includegraphics[keepaspectratio, width=0.3\linewidth]{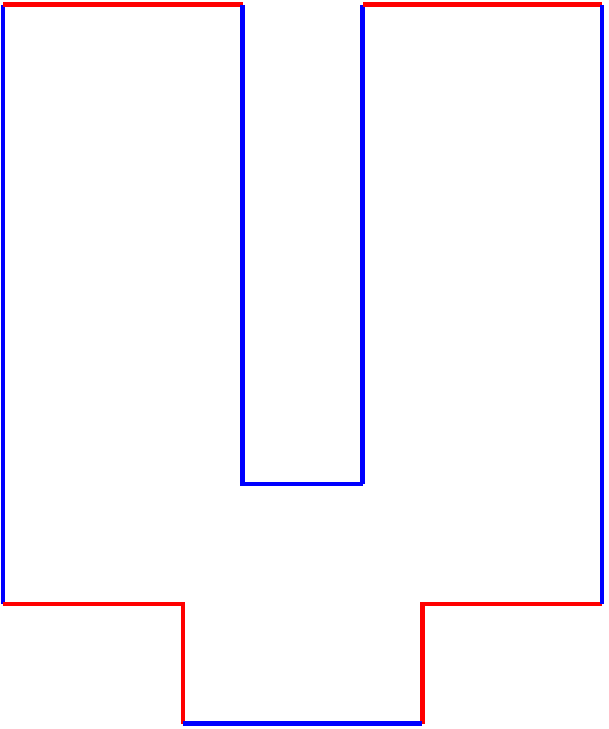}
    }
    \hspace{1em}
    \subfloat[][The two protruding parts at the top have been reduced downward. The densely packed surface codes have finished shifting downward.]{
    \includegraphics[keepaspectratio, width=0.3\linewidth]{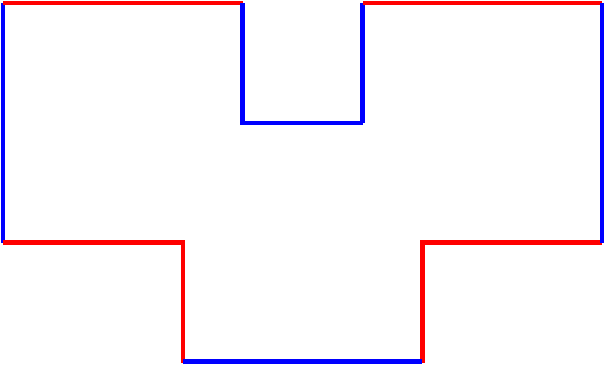}
    }
    \caption{Row-shifting movement in the densely packed surface code.}
\label{fig:movement_of_dense}
\end{figure*}

We present the deformation procedure for the movement of an entire row of the densely packed surface codes. Fig.~\ref{fig:movement_of_dense} illustrates the deformation procedure in a condensed manner. The red lines correspond to $X$ boundaries and the blue lines correspond to $Z$ boundaries.
The physical operations required for the deformation, such as initialization of the state of physical qubits, stabilizer measurements, and measurement of physical qubits, are the same as those described in the Appendix~\ref{sec:app_deform}.

Fig.~\ref{fig:movement_of_dense}(a) shows the densely packed surface code before the movement. The physical qubits located below the bottom $Z$ boundary of the densely packed surface code are initialized in the $\ket{0}$ state, and stabilizer measurements are performed, including the newly introduced stabilizer generators. After $d$ rounds of stabilizer measurements, the bottom of the densely packed surface code is extended, as shown in Fig.~\ref{fig:movement_of_dense}(b). This operation corresponds to the extension of the logical $Z$ operator associated with the lower codeword in the densely packed surface code.

Next, we initialize the physical qubits to the $\ket{+}$ state along the $X$ boundaries connecting the top and bottom regions of the densely packed surface code. We perform stabilizer measurements, including new stabilizer generators. After $d$ rounds of stabilizer measurements, we obtain the expanded surface codes shown in Fig.~\ref{fig:movement_of_dense}(c). This operation corresponds to extending the logical $X$ operators of the upper two codewords downward.

We measure the physical qubits in the $Z$ basis to shorten the logical $Z$ operator of the lower codeword, obtaining the state shown in Fig.~\ref{fig:movement_of_dense}(d). Similarly, we measure the physical qubits in the $X$ basis to shorten the local operators of the upper two codewords. Fig.~\ref{fig:movement_of_dense}(e) shows the result, and the entire densely packed surface code has moved downward from the state shown in Fig.~\ref{fig:movement_of_dense}(a). From the above, one can move the entire densely packed row downward with $2d$ rounds plus two parallel physical qubit measurement steps.

An alternative way to move the densely packed surface code is to repeatedly fuse and separate patches. In this approach, the lower patches are separated and shifted downward, followed by re-fusing the upper patches. However, this method requires an additional $d$ rounds compared to the procedure in Fig.~\ref{fig:movement_of_dense}, and thus we adopt the latter.

\clearpage

\end{document}